\documentclass[a4paper,12pt]{article}      
\usepackage{graphicx}
\usepackage{amsmath}
\usepackage{amsfonts}

\begin{document}

\begin{flushright}			
       CHIBA-EP-124-REV\\
       HEP-TH/0012043\\
       \tiny{Compiled \today}
\end{flushright}
\begin{center}

\renewcommand{\thefootnote}{\fnsymbol{footnote}}

{
\Large \sffamily \bfseries%
  {%
	Consistency of the Hybrid Regularization with\\
	Higher Covariant Derivative and\\
	Infinitely Many Pauli-Villars\\
  }
}

\vspace{18pt}
{\large \sffamily \bfseries {Koh-ichi Nittoh}}
      \footnote{E-mail: \texttt{tea@cuphd.nd.chiba-u.ac.jp}}
\\
{\small\it
Center for Frontier Electronics and Photonics, Chiba University,\\
1-33 Yayoi-cho, Inage-ku, Chiba 263-8522, Japan}\\
\end{center}\vspace{18pt}
\begin{abstract}
We study the regularization and renormalization of the Yang-Mills theory
in the framework of the manifestly invariant formalism,
which consists of a higher covariant derivative
with an infinitely many Pauli-Villars fields.
Unphysical logarithmic divergence,
which is the problematic point on the Slavnov's method,
does not appear in our scheme,
and the well-known vale of the renormalization group functions are derived.
The cancellation mechanism of the quadratic divergence
is also demonstrated
by calculating the vacuum polarization tensor
of the order of $\Lambda^0$ and $\Lambda^{-4}$.
These results are the evidence that
our method is valid for intrinsically divergent theories
and is expected to be available for the theory which contains the quantity
depending on the space-time dimensions, like supersymmetric gauge theories.
\end{abstract}

\renewcommand{\thefootnote}{\arabic{footnote}}
\setcounter{footnote}{0}

\section{Introduction}

Manifestly gauge invariant regularization is useful
for the gauge invariant renormalization,
but there is not so many invariant regularization
for non-Abelian gauge theories.
The dimensional regularization is the most popular method,
but is not manifestly invariant method for the theory
which contains quantities depending on the space-time dimension.
%
The chiral gauge theory is a good example of the theory
which is not regularized by the dimensional regularization
because the symmetry depending on $\gamma_5$ is relevant.
%
Frolov and Slavnov proposed an invariant regularization method for this theory
introducing an infinite number of Pauli-Villars (PV) fields%
~\cite{Frolov-Slavnov93,Frolov-Slavnov94}.

There is a similar problem 
in the three-dimensional Chern-Simons (CS) gauge theory,
where the anti-symmetric symbol $\epsilon_{\mu\nu\rho}$
is defined by the space-time dimension.
%
For this theory, we developed a parity-invariant PV regularization
introducing an infinite number of PV fields%
~\cite{Nittoh-Ebihara98MPLA,Nittoh99phd}.
Using with the higher covariant derivative (HCD) method,
we can construct a hybrid regularization of the HCD and the PV
without breaking a parity invariance.
By this regularization,
we can classify the universality class of the CS coupling shift
in the framework of the hybrid regularization
~\cite{Nittoh-Ebihara98MPLA,Nittoh99phd}.

Another example of the theory depending on the space-time dimension
is the supersymmetric (SUSY) gauge theory.
For this theory,
it is believed that a manifestly invariant regularization exists
but there is few method preserving both the gauge symmetry and supersymmetry%
~\cite{Jack-Jones97}.
Since the PV regularization and the HCD regularization are defined 
independently of the space-time dimension,
the hybrid regularization is expected to be one of the candidates
for the invariant regularization for SUSY.
It is indeed available for the three-dimensional SUSY gauge theory
and gives the same value of the CS coupling shift 
which is predicted by Kao et al.~\cite{Kao-Lee-Lee95}
in the SUSY Yang-Mills-Chern-Simons (YMCS) theory%
~\cite{Koiwa98}.

When our method is extended to the four-dimensional SUSY  gauge theory,
it is not obvious whether our regularization works properly
because the intrinsically divergence appears in four-dimensions
though the three-dimensional theory is finite.
%
So we have to confirm that
our method is available not only the finite theory
but also the divergent theory
to construct a regularization scheme base on the hybrid regularization
that works properly whenever the dimensional regularization fails.
For this purpose,
we consider the regularization and renormalization
of the four-dimensional Yang-Mills (YM) theory in this paper,
which is the simplest model of the divergent theory.

%
The hybrid regularization was originally proposed by Slavnov in 1970s%
~\cite{Slavnov72,Slavnov77,Faddeev-Slavnov91}.
When the HCD method is applied to the non-Abelian gauge theory,
some one-loop diagrams are left unregularized
because the HCD term renders the gauge propagator less divergent
but vertices more divergent.
To regularize these remaining diagrams,
other regulator must be introduced.
Slavnov employed the PV method for the regulator of the remaining diagrams.

Though the theory was formally regularized by his regularization,
it was pointed out that his method leads wrong value of 
the renormalization group (RG) $\beta$- and $\gamma$-function
for the four-dimensional YM theory%
~\cite{Martin-Ruiz95}.
Moreover, the unitarity breaking was confirmed
when his method is applied to the quantum chromodynamics%
~\cite{Leon-Martin-Ruiz95}.
%
These problems were occurred
by the unphysical non-local logarithmic radiative corrections
from the PV determinant.
Namely, the Slavnov's PV field was not the complete regulator
for the remaining one-loop divergence.

To overcome this problem, two modifications have been proposed.
One is to use the dimensional regularization
instead of the Slavnov's PV regularization%
~\cite{Martin-Ruiz95PL}.
This scheme does not lead the unphysical logarithmic corrections,
but is not suit for our aim to apply to the theory
which is not regularized by the dimensional regularization.
%
Another proposal is to modify the Slavnov's PV fields
not to lead the unphysical corrections.
Introducing `gauge-fixing parameter' for the PV fields,
this task is accomplished with preserving the gauge symmetry%
~\cite{Asorey-Falceto95}.

On these two proposals,
the extra regularization so-called `pre-regulator'
is needed to evaluate the divergence of the diagrams.
Since it is inserted as a partial regulator,
the regularization might be inconsistent in the scheme.
If we use the dimensional regularization as the pre-regulator
in the modified hybrid regularization,
the scheme is not suit for the regularization of the theory
depending on the space-time dimension.
The extra regulator is sometimes unwanted procedure for our aim
so we have to develop a procedure without it.

On the other hand,
the four-dimensional YM theory leads up to the quadratic divergence.
Since the quadratic divergence is not ignored
in the PV type of regularization
though it does not appear in the dimensional regularization essentially,
we also confirm whether it is canceled or not
in the hybrid regularization scheme.
%
The cancellation, however, is not verified so far,
though the logarithmic divergence is shown
to agree with the physical divergence
derived from the other regularization scheme~\cite{Pronin-Stepanyantz97PL}.

The reason why
the cancellation does not confirmed
and is not problematic in references~\cite{Martin-Ruiz95,Leon-Martin-Ruiz95}
is because they use the dimensional regularization as the `pre-regulator'
in addition to the fact that the quadratic divergence has no physical meanings.
In their scheme,
the quadratic divergence does not appear in principle
though the HCD term leads the non-trivial contribution
of quadratically divergent diagrams to the effective action.
%
So the explicit cancellation of the quadratic divergence must be confirmed
in our regularization scheme
for the complete regularization.

The organization of this paper is follows.
%
In Section~\ref{sec:the regularization method},
we apply our regularization scheme to the four-dimensional YM theory
writing the explicit form of the action.
It is confirmed that
all the divergences are regularized by HCD terms
except some diagrams at one-loop level by the superficial degree of divergence.
A minor modification for the HCD term is given
for the complete cancellation of the quadratic divergence.
To treat the unregularized one-loop diagrams,
the infinitely many PV fields are introduced
in a similar way of the CS gauge theory.
We check whether the PV fields really cancel the divergence
by an explicit calculation of one-loop diagrams
in the following sections.
%
In Section~\ref{sec:one-loop contributions},
we give a discussion for the contributions independent of $\Lambda$.
Though each diagram contains up to the quadratic divergence,
almost the divergence cancels out in total
and only the logarithmic divergence remains.
The renormalization is also given
to treat the logarithmic divergence
following the usual procedure.
After that, our regularization scheme works properly
and gives the correct RG $\beta$- and $\gamma$-function
in this calculation.
We also show that any explicit pre-regulator is not necessary
to see the cancellation of the quadratic divergence.
%
In Section~\ref{sec:one-loop contributions in finite},
the one-loop contributions of $\Lambda^{-4}$ order
are calculated.
It is show that
the cancellation mechanism of the divergence is the same as
one of the $\Lambda^0$ order.
We also show that
the cancellation mechanism works
only when we use the modified HCD term
in Section~\ref{sec:the regularization method}.
%
Conclusions and discussions are given in the last section
and some useful formulae are given in Appendices
which are employed in explicit calculations of one-loop diagrams
in Section~\ref{sec:one-loop contributions}%
~and~\ref{sec:one-loop contributions in finite}.

\section{The Regularization Method}
\label{sec:the regularization method}

We consider the hybrid regularization of the YM theory
in this section.
%
The hybrid regularization consists of the following two steps.
First we introduce HCD terms.
They improve the behavior of propagators at large momentum,
rendering the theory less divergent
at the cost of 
irrelevant vertices.
The theory is reduced to superrenormalizable one
which has just a finite number of divergent loops.
As see later,
all the diagrams except one-, two-, three- and four-point functions
at one-loop level are convergent with a suitable choice of these terms.
Secondly, we deal with 
unregularized diagrams by a PV type of regularization.
Since we are considering the gauge invariant regularization,
the PV regulator
must be constructed gauge invariant form
and never lead any unphysical divergence.

The generating functional regularized by the hybrid regularization
is written by
\begin{equation}
Z=
   \int \mathcal{D}A_\mu \mathcal{D}b
        \mathcal{D}\overline c \mathcal{D}c
\,\exp[-S_\Lambda]
\prod_j
  \det{}^{-\frac{\alpha_j}{2}}\mathbf{A}_j
\prod_i
  \det{}^{\gamma_i}\mathbf{C}_i,
\label{eq:generating functional}
\end{equation}
%
where $S_\Lambda$ is an action regularized by HCD terms,
$\det{}^{-\frac{\alpha_j}{2}}\mathbf{A}_j$ and
$\det{}^{\gamma_i}\mathbf{C}_i$ are PV determinants
for the gauge and ghost respectively.
In the following,
we consider the regularization in detail
and write the explicit form of the HCD terms and PV determinants.

In four-dimensional Euclidean space-time,
the YM theory is given by the action
\begin{equation}
S=S_\mathrm{YM}+S_\mathrm{GF},
\label{eq:classical action}
\end{equation}
where 
\begin{align}
S_\mathrm{YM}&=
\frac{1}{ 4}\int \mathrm{d}^4x\,
F_{\mu\nu}^a F^{\mu\nu}{}^a,
\label{eq:YM action}\\
S_\mathrm{GF}&=
\int \mathrm{d}^4x\,
 \frac{\xi_0}{ 2}b^a b^a
 -b^a (\partial^\mu A_\mu)^a
 +\overline c ^a (\partial_\mu D^\mu c)^a,
\label{eq:GF action for YM}
\end{align}
with the field strength 
$
F_{\mu\nu}^a = 
\partial_\mu A_\nu^a - \partial_\nu A_\mu^a
+g f^{abc}A_\mu^b A_\nu^c
$
and the covariant derivative
$
D_\mu^{ac} =
\delta^{ac}\partial_\mu
+ gf^{abc}A_\mu^b.
$
Here $A_\mu^a$, $c^a$, $\overline c ^a$ and $b^a$
denote the gauge field, ghost, anti-ghost
and auxiliary field respectively,
$\xi_0$ is the gauge-fixing parameter 
and $f^{abc}$ is the structure constant of the gauge group SU($N$).
We abbreviate the color index in the following discussions.

  \subsection{Higher covariant derivative method}
  \label{sec:Higher covariant derivative method}

The basic idea of HCD method is to regularize the diagrams
by improving the convergence of propagators
with higher derivative terms.
When we choose $\Lambda$ as a cutoff parameter,
the most general form of the HCD action is given by
\begin{equation}
S_\mathrm{HCD} =
\frac{1}{ 4 \Lambda^{2n}}\int \mathrm{d}^4x\,
D^nF_{\mu\nu}D^nF^{\mu\nu}.
\label{eq:general HCD action}
\end{equation}
Though the original propagator derived from \eqref{eq:classical action}
behaves $\sim p^{-2}$ at large momentum $p$,
the large momentum behavior of the transverse part of the propagator
is modified to $\sim p^{-2-2n}$
after the insertion of \eqref{eq:general HCD action},
but the longitudinal part is not.
So the whole of the propagator still behaves
$\sim p^{-2}$ at large momentum $p$.
To improve the convergence of the longitudinal part,
we introduce a \textit{higher derivative} term $H$
to the gauge-fixing action as follows~\cite{NittohEP118};
\begin{equation}
S_\mathrm{GF}^{H} =
\int \mathrm{d}^4x\,
 \frac{\xi_0 }{ 2}b^2
 -b H \partial^\mu A_\mu
 +\overline c H \partial_\mu D^\mu c.
\label{eq:GF with HD action}
\end{equation}
$H$ is a dimensionless function of $\partial^2/\Lambda^2$
and must contains a higher term than $\partial^n/\Lambda^n$
to ensure the large momentum behavior of the propagator.

The most distinctive point of our regularized action is
that the \textit{higher derivative} term for the ghost is
introduced in \eqref{eq:GF with HD action},
which is necessary for the BRST invariance on our method.
There is another choice of the regularized action
according to the references~\cite{Martin-Ruiz95,Martin-Ruiz95PL}
which does not need any higher derivative term for the ghost.
It seems simpler than our method and we prefer to use their action,
but as we see later,
the quadratic divergence is not completely canceled
among the $\Lambda$-dependent terms~\cite{NittohEP118}.
We will come back to this problem in more detail later
in Section~\ref{sec:Inconsistent higher derivative action}.

So the regularized action in \eqref{eq:generating functional} is given by
\begin{equation}
S_\Lambda = S_\mathrm{YM}+S_\mathrm{HCD}+S_\mathrm{GF}^{H},
\label{eq:regularized action}
\end{equation}
%
%
and invariant under the BRST transformations
\begin{align}
\delta_\mathrm{B} A_\mu &= (D_\mu c),&
\delta_\mathrm{B} b     &= 0,&
\delta_\mathrm{B} c     &= - c\times c,&
\delta_\mathrm{B} \overline c &= b.
\label{eq:BRST transformations}
\end{align}
Here $\psi \times \phi$ means $gf^{abc} \psi^b \phi^c$
and then the BRST operator $\delta_\mathrm{B}$ satisfies
the usual nilpotency $\delta_\mathrm{B}^2=0$.

The superficial degree of divergence is calculated at
\begin{equation}
\omega = 4 - 2n (L-1)
- E_A - \left(\frac{n}{2}+1 \right) E_c,
\label{eq:degree of divergence}
\end{equation}
where $L$, $E_A$ and $E_c$ are the number of
loops, external line of the gauge 
and external line of the ghost, respectively.
For all the diagrams higher than two-loop ($L\ge2$),
$n \ge 2$ always gives negative $\omega$.
This means that
we may remove the higher loops by a suitable choice of $n$.
%
%
On the other hand 
for one-loop ($L=1$),
$\omega$ is not always negative by any $n$.
The most economical choice is $n=2$ and we adopt it.

The  explicit form of $H$ is determined
by the behavior of the gauge propagator
which is obtained as follows:
\begin{equation}
\frac{\Lambda^4 }{ p^4 (p^4 + \Lambda^4)}
(p^2 \delta_{\mu\nu} - p_\mu p_\nu)
+
\frac{\xi_0 }{ p^4 H^2({p^2 }/{ \Lambda^2})}
p_\mu p_\nu.
\label{eq:gauge propagator}
\end{equation}
The first term has the order of the momentum degree of $-6$,
the second term must be the same degree or less
to ensure the convergence of the diagrams
except one-, two-, three- and four-point functions at one-loop level;
$H^2$ behaves $\sim p^4$ at large $p$.
When we naively remove the higher derivative terms
taking the limit of $\Lambda \rightarrow \infty$,
the propagator must recover familiar one;
$H^2$ converges to unity.
So the simplest form of $H^2$ in momentum space is%
\begin{equation}
H^2
\left(\frac{p^2 }{ \Lambda^2}\right)
=
1 + \frac{p^4 }{\Lambda^4}.
\label{eq:explicit H2}
\end{equation}
It reduces the propagator to very simple form
in which the first and the second term of (\ref{eq:gauge propagator})
has the same denominator.
Especially in Feynman gauge ($\xi_0 =1$)
the propagator is reduced to
\begin{equation}
\frac{\Lambda^4 }{ p^2 (p^4 + \Lambda^4)}
\delta_{\mu\nu},
\end{equation}
which renders diagrams simpler
so we choose the Feynman gauge in the following sections.

  \subsection{Regularization of one-loop divergence}

All the diagrams are regularized except one-, two-, three- and four-point
functions at one-loop level.
We need an extra regularization to regularize the remaining one-loop diagrams.
It was shown in reference~\cite{Faddeev-Slavnov91}
that the PV regularization does not break gauge invariance
at one-loop level
and therefore can be used to complete the regularization of the theory.

%
The PV determinant of \eqref{eq:generating functional} is introduced
such as
\begin{equation}
\prod_j
  \det{}^{-\frac{\alpha_j}{2}}\mathbf{A}_j
=
\prod_{j=1}^\infty
 \det{}^{-\frac{\alpha_{+j}}{2}}\mathbf{A}_{+j}
 \det{}^{-\frac{\alpha_{-j}}{2}}\mathbf{A}_{-j}.
\label{eq:PV determinant}
\end{equation}
%
The basic idea of our PV regularization scheme is
to regularize the theory by the pair of two PV determinants
$\det{}^{-\frac{\alpha_{+j}}{2}}\mathbf{A}_{+j}$ and
$\det{}^{-\frac{\alpha_{-j}}{2}}\mathbf{A}_{-j}$.
These determinants are the generalization of
ones in three dimensional CS gauge theory:
where
a pair is needed to make a parity-invariant PV regulator,
because a part of the action breaks the invariance%
~\cite{Nittoh-Ebihara98MPLA}.
The same situation also arises
in the regularization of the chiral gauge theory%
~\cite{Frolov-Slavnov93,Frolov-Slavnov94},
the regularization by the pair is regarded as a more general method
than the usual PV regulator.
So the YM theory ought to be regularized by the similar way of the pair.

An infinite number 
is needed to use such PV pairs.
Introducing one pair corresponds to subtracting double the divergence,
so we have to remedy the over subtraction
by an insertion of another pair of opposite statistics.
Then, to remedy the over addition
we have to introduce the third pair.
Such steps are repeated alternately until the divergence is removed.
The divergence does not converged by finite steps but infinite,
so we must introduce an infinite number of the PV pairs.

Such steps correspond to introducing
fermionic PV fields $(\alpha_j=-1)$
and bosonic PV fields $(\alpha_j=+1)$ alternately.
%
Following references~\cite{Nittoh-Ebihara98MPLA,Nittoh99phd},
we take the PV conditions for the gauge field such as
\begin{align}
M_{\pm j}&=M|j|,&
\alpha_{\pm j}&=(-1)^{|j|}.
\label{eq:PV conditions}
\end{align}
$M_j$ denotes the mass parameter of the PV field $A_j{}_\mu$,
and the determinant is represented as
\begin{equation}
\det{}^{-\frac{\alpha_j}{2}}\mathbf{A}_j =
\int \mathcal{D}A_j{}_\mu \mathcal{D}b_j
\exp[-S_{M_j}-S_{b_j}^H],
\label{eq:det A}
\end{equation}
where
\begin{align}
S_{M_j}&=
\frac{1}{ 2}\int \mathrm{d}^4x \mathrm{d}^4y
A_j{}_\mu(x)
\left[
 \frac{\delta^2 S_\Lambda }{ \delta A_\mu(x) \delta A_\nu(y)}
 -M_j^2 g^{\mu\nu} \delta(x-y)
\right]
A_j{}_\nu(y),
\label{eq:PV for gauge action}
\\
S_{b_j}^H&=
\int \mathrm{d}^4x
\left[
 \frac{\xi_j }{ 2}
 b_j b_j
 -b_j {\Tilde H} D^\mu A_j{}_\mu
\right].
\label{eq:auxiliary field for PV action}
\end{align}
We do not take the summation with the index $j$
in \eqref{eq:PV for gauge action}
and \eqref{eq:auxiliary field for PV action}.
$b_j$ is an auxiliary field for $A_j{}_\mu$
and $\xi_j$ a `gauge-fixing parameter' for the PV field.
In usual PV regularization,
the `gauge-fixing parameter' for the PV field is ordinary chosen
in Landau gauge $\xi_j =0$.
In this gauge, however,
PV determinants do not converge formally to a constant
and then the anomalous divergence appears~\cite{Asorey-Falceto95}.
This anomalous divergence contributes to the renormalization
and gives the wrong RG $\beta$- and $\gamma$-function%
~\cite{Martin-Ruiz95,Asorey-Falceto95}.
To resolve such a problem,
we have to introduce $\xi_j (\ne 0)$ 
in \eqref{eq:auxiliary field for PV action}.

%
${\Tilde H} = {\Tilde H}(D^2/\Lambda^2)$
is a higher \textit{covariant} derivative term
which has the same effect as $H(\partial^2 / \Lambda^2)$
and satisfies the following two conditions.
First,
it must behave $\sim p^4$ 
to improve the behavior of the propagator at large momentum
and converges to unity at the limit $\Lambda \rightarrow \infty$
to recover the usual PV fields.
Secondly, it must be invariant under the BRST transformations
which come from
the change of the integration variables
$\phi_j\rightarrow \phi_j +\theta (\delta_\mathrm{B} \phi_j)$,
where $\theta$ is an anti-commuting parameter
and $\phi_j$ denotes a PV field,
then $\delta_\mathrm{B} \phi_j$ is given by
\begin{align}
\delta_\mathrm{B}
   A_j{}_\mu &= A_j{}_\mu \times c,&
\delta_\mathrm{B}
   b_j &= b_j \times c,
\label{eq:BRST transformations for PV 1}
\end{align}
On these conditions,
${\Tilde H}(D^2/\Lambda^2)$ must be the polynomial of the
\textit{covariant derivative}
and the simplest form is given by
\begin{equation}
{\Tilde H}^2 \left(\frac{D^2}{\Lambda^2}\right)
 =1+\frac{D^4}{\Lambda^4}.
\end{equation}
Consequently the PV determinant \eqref{eq:det A}
is constructed to preserve the BRST invariance.

Similarly,
the PV determinant for the ghost is written
\begin{equation}
\prod_i
  \det{}^{\gamma_i}\mathbf{C}_i
=
\prod_{i=1}^\infty
 \det{}^{\gamma_{+i}}\mathbf{C}_{+i}
 \det{}^{\gamma_{-i}}\mathbf{C}_{-i},
\end{equation}
\begin{equation}
\det{}^{\gamma_i}\mathbf{C}_i =
\int \mathcal{D}\overline c_i\mathcal{D}c_i
\exp
\left[
  -\int \mathrm{d}^4x
  \left(
   \overline c_i {\Tilde H} D_\mu D^\mu c_i
   -m_i^2 \overline c_i c_i
  \right)
\right],
\label{eq:det C}
\end{equation}
where we do not take the summation with the index $i$ again.
$\overline c_i$ and $c_i$ are PV fields for the ghost and anti-ghost
of mass $m_i$.
We can treat these PV fields under the PV conditions
$m_{\pm i}=m|i|$ and $\gamma_{\pm i}=(-1)^{|i|}$.
The HCD term for the PV field is inserted as the function
$\Tilde H$ to preserve the BRST invariance of
\begin{align}
\delta_\mathrm{B}
   c_i &= c_i \times c,&
\delta_\mathrm{B}
   \overline c_i &= \overline c_i \times c.
\label{eq:BRST transformations for PV 2}
\end{align}
This HCD term is necessary for the cancellation of the quadratic divergence
when the mass term is simply introduced in the usual form
like $m_i^2 \overline c_i c_i$,
as we see in Section~\ref{sec:one-loop contributions in finite}.

In the following sections,
we confirm that such PV fields completely regularize the theory
by an explicit evaluation of the one-loop diagrams.
In this calculation, one of the most important procedure is 
to summate an infinite number of the PV diagrams
and derive a convergent function from them.
Since this procedure is carried out before the momentum integration,
all the parameters independent of the indices $i$ and $j$
must be chosen to be the same.
So we assign the same momentum parameter to the internal line of each diagram
when all the one-loop diagrams are drawn in the next section.

  \subsection{Feynman rules}

The regularized action is decomposed
into the kinetic part $K$ and the vertex part $V$ as follows%
~\cite{Pronin-Stepanyantz97PL,Pronin-Stepanyantz97NP}:
\begin{equation}
\int \mathrm{d}^4x \Psi (x) (K + V + M^2) \Phi (x),
\end{equation}
where $\Psi(x)$ and $\Phi(x)$ denote arbitrary fields
and $M$ their mass parameter.
Since $K$ and $V$ consist of the original part from the YM term
(denoting with suffix `0')
and $\Lambda$-dependent part from the HCD term
(with suffix `$\Lambda$'),
they are decomposed into
\begin{align}
K&=K_0+\frac{1}{\Lambda^4}K_\Lambda,&
V&=V_0+\frac{1}{\Lambda^4}V_\Lambda.
\end{align}
Under this decomposition,
the propagators are written in the form
\begin{equation}
\frac{1}{K+M^2}
=\frac{1}{K_0+M^2}
 \left(
   1-\frac{K_\Lambda}{K_0+M^2}\Lambda^{-4}+O(\Lambda^{-8})
 \right).
\end{equation}
So the Feynman rules are written by the order of $\Lambda^{-1}$,
the quantum corrections are calculated order by order.
The Feynman rules are listed in Appendix~\ref{appendix:Feynman Rules}.

\section{One-Loop Contributions Independent of $\Lambda$}
\label{sec:one-loop contributions}

Now we check whether the PV fields cancel the unregularized divergence
by an explicit calculation of the one-loop contributions.
First we calculate the contribution which does not depend on $\Lambda$.
%

%

All the diagrams are manipulated under the following three rules.
\begin{enumerate}

\item
Take the same assignment for the internal momentum
among graphically the same form.

\item
Take the infinite sum of PV diagrams
under the PV conditions \eqref{eq:PV conditions}
adding a `virtual' PV diagram constructed by
taking the massless limit of the PV field.

\item
Subtract the `virtual' diagram to ensure the total contribution
as a diagram from original fields

\end{enumerate}
As we mentioned in the previous section,
Rule~1 is necessary to find a convergent function easily
after the infinite sum of the diagrams.
Rule~2~and~3 means to divide the contribution of the PV determinants
into two parts,
the part of the infinite sum and of the massless term.

In the `virtual' PV diagram,
the `zeroth' field like `$A_0{}_\mu$' runs as the internal propagator.
We call such a diagram as `zeroth' diagram in the below.
The same contribution must be subtracted
to maintain the total contribution.
This procedure is realized by the following equation:
\begin{equation}
\prod_{j=1}^\infty
 \det{}^{-\frac{\alpha_j}{2}}\mathbf{A}_{+j}
 \det{}^{-\frac{\alpha_{-j}}{2}}\mathbf{A}_{-j}
=
{\prod_{j=-\infty}^{\infty}
\det{}^{-\frac{\alpha_j}{2}}\mathbf{A}_j}
/
{\det{}^{-\frac{\alpha_0}{2}}\mathbf{A}_0},
\label{eq:determinant of infinite sum}
\end{equation}
and so on.
The series is defined to give an uniquely convergent function
when we go to the r.h.s. of this equation.

In the below,
we denote the contribution from the infinite sum by `massive' contribution
and the other by `massless' contribution.
The contribution from the diagrams of original fields
and of massless zeroth PV fields belongs to the latter.
This decomposition of the contribution clarifies
the cancellation of the quadratic divergence
as we see in the below.

  \subsection{Vacuum polarization tensor}

\begin{figure}[t]
\begin{center}
\includegraphics
{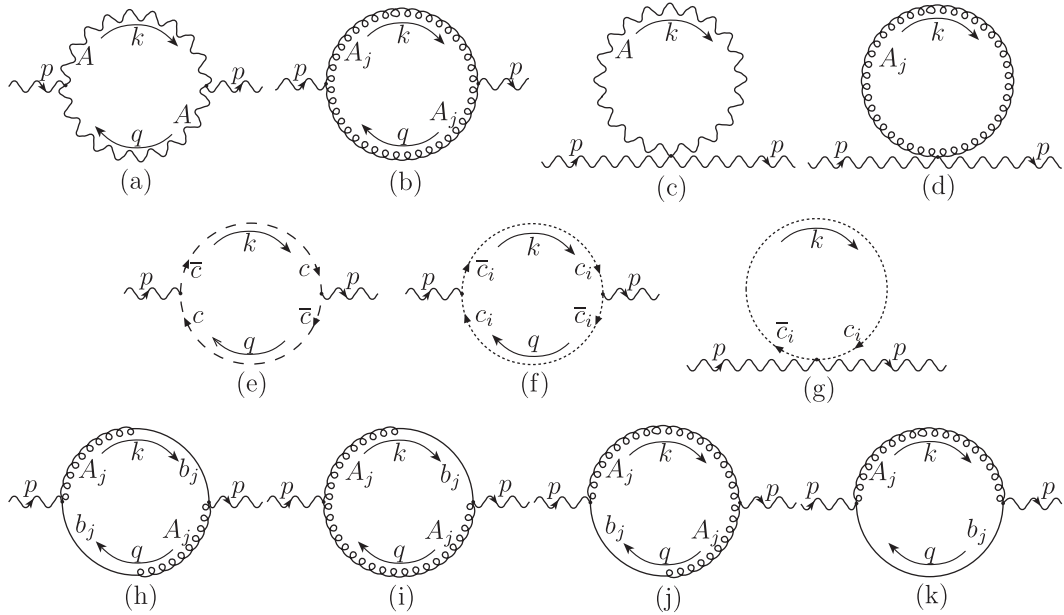}
\caption
{All the diagrams contribute to the vacuum polarization tensor
 at $\Lambda^0$ order.
 The wavy line means the gauge field $A$,
 the curly line the PV field $A_j$,
 and the straight line the auxiliary field $b_j$.
 The rough- and fine-dotted line mean
 ghost $c$ and PV for ghost $c_i$, respectively.
 We use the same assignment of internal momenta, where $q=k-p$.
}
\label{fig:loops}
\end{center}
\end{figure}

All the diagrams contributing to the vacuum polarization tensor
at $\Lambda^0$ order
are listed in Figure~\ref{fig:loops}.
We denote the quantum correction from the diagram (b)
containing the internal field $A_j{}_\mu$
by $\overset{_\mathrm{(b)}}\Pi_j{}_{\mu\nu}(p)$ and so on.
The Feynman gauge $\xi_0 = 1$ is chosen for its simplicity
in which the propagator is reduced to the simplest form
as mentioned in the last section.
The gauge fixing parameter for the PV field is taken as the same value,
$\xi_j = 1$, for the same reason.
All the diagrams are divided into three groups
and the quantum corrections are calculated in each group.

    \subsubsection{Gauge field type (A-type) diagrams}
The diagrams (a), (b), (c) and (d) in Figure~\ref{fig:loops}
belong to this group.
First we consider the diagram (b)
that contains only $A_j{}_\mu$ field in the internal line.
To take the infinite sum following \eqref{eq:determinant of infinite sum},
the diagram of `$A_0{}_\mu$' must be introduced to our calculation.
Notice the diagram (a) has the same structure with the zeroth diagram of (b),
both the diagrams are identifiable as 
$
\overset{_\mathrm{(a)}}\Pi{}_{\mu\nu}(p)\equiv
\overset{_\mathrm{(b)}}\Pi_0{}_{\mu\nu}(p).
$
The infinite sum is taken without any extra virtual diagram.
All the contribution from these diagrams are written
in the form of the infinite sum
which is denoted by `massive' contribution as follows:
\begin{multline}
\overset{_\mathrm{(a)}}\Pi{}_{\mu\nu}(p)
+
\sum_{\substack{j=-\infty \\ j\ne 0}}^\infty
\alpha_j\overset{_\mathrm{(b)}}\Pi_j{}_{\mu\nu}(p)
\bigg|^{\Lambda^0}
=
\sum_{j=-\infty}^\infty 
(-1)^j\overset{_\mathrm{(b)}}\Pi_j{}_{\mu\nu}(p)
\bigg|^{\Lambda^0}
\\
\shoveleft{
=\frac{g^2 c_v}{ 32 \pi^2}
\left[
\frac{3M^2 }{ 10}C_2\delta_{\mu\nu}+
\left(
 \frac{22}{ 3} \ln\left(\frac{\pi p }{ 2M}\right)
 -\frac{61 }{ 9} - \frac{\pi^2 p^2 }{ 9M^2}
\right)p_\mu p_\nu
\right.
}\\
\left.
-
\left(
 \frac{19}{ 3} \ln\left(\frac{\pi p }{ 2M}\right)
 -\frac{49 }{ 9} - \frac{37 \pi^2 p^2 }{ 360 M^2}
\right)p^2 \delta_{\mu\nu}
+O\left(M^{-4}\right)
\right],
\label{eq:summation of (a) and (b)}
\end{multline}
where we use the PV conditions \eqref{eq:PV conditions}
and the formula \eqref{eq:result of infinite sum}.
Under these conditions
all the summations are calculated
in the same manner with the matter field in the chiral gauge theory%
~\cite{Frolov-Slavnov93,Frolov-Slavnov94}
using the formula
$\sum_{j=-\infty}^{\infty}
\frac{(-1)^j }{ A^2 + j^2}
 =\frac{\pi }{ A \sinh \pi A}$.
Then the massive contribution converges in the finite $M$
as is shown in \eqref{eq:result of infinite sum}
in Appendix~\ref{appendix:calculations of infinite sum}.
The first term of the second line, multiplied by $C_2$,
expresses the quadratic divergence at $M\rightarrow \infty$.
$C_2$ is a dimensionless constant which originates from
the integration with $X=p^2/M^2$.

%
In the same way,
we calculate the corrections from the diagrams (c) and (d)
identifying
$
\overset{_\mathrm{(c)}}\Pi{}_{\mu\nu}(p) \equiv
\overset{_\mathrm{(d)}}\Pi_0{}_{\mu\nu}(p),
$
\begin{align}
\overset{_\mathrm{(c)}}\Pi{}_{\mu\nu}(p)
+\sum_{\substack{j=-\infty \\ j\ne 0}}^\infty
\alpha_j\overset{_\mathrm{(d)}}\Pi_j{}_{\mu\nu}(p)
=
-\frac{g^2c_v}{ 32 \pi^2}
\frac{M^2}{5}C_2\delta_{\mu\nu}
+O\left(M^{-4}\right).
\label{eq:summation of (c) and (d)}
\end{align}
This contribution of \eqref{eq:summation of (c) and (d)}
does not lead the logarithmic divergence,
but give the quadratic divergence as well as
\eqref{eq:summation of (a) and (b)}.

    \subsubsection{Auxiliary field type (B-type) diagrams}
For the diagrams (h), (i), (j) and (k),
the zeroth diagram is needed to take the infinite sum of these diagrams.
There is, however, no diagram which is identified as such a zeroth diagram
for this group.
So we have to add the virtual diagrams for (h), (i), (j) and (k)
where $b_0$ or $A_0$ field runs as an internal propagator
instead of $b_j$ or $A_j$.
Then the contribution from the diagram (h) is calculated
\begin{equation}
\sum_{\substack{j=-\infty \\ j\ne 0}}^\infty
\alpha_j\overset{_\mathrm{(h)}}\Pi_j{}_{\mu\nu}(p)
=
\sum_{j=-\infty}^\infty
(-1)^j\overset{_\mathrm{(h)}}\Pi_j{}_{\mu\nu}(p)-
\overset{_\mathrm{(h)}}\Pi_0{}_{\mu\nu}(p).
\label{eq:cure for no massless diagrams}
\end{equation}
The second term of the r.h.s. is a counter term for the zeroth diagram of (h)
introduced to extract the massive contribution of the first term.
Such a counter term is classified into `massless' contribution.
We take the same care for the other diagrams (i), (j) and (k).
\footnote
{The compensation term for the diagram (k),
 $\overset{_\mathrm{(k)}}\Pi_0{}_{\mu\nu}(p)$, is zero
 because the mass parameter $M_j$ is multiplied
 in the numerator of the integrand all over
 when we write the contribution explicitly in the integral form.
 But we formally write it in the following discussions.}
Then the total of massive contributions is
\begin{multline}
\sum_{j=-\infty}^\infty
(-1)^j
\left(
   \overset{_\mathrm{(h)}}\Pi_j{}_{\mu\nu}(p)
  +\overset{_\mathrm{(i)}}\Pi_j{}_{\mu\nu}(p)
  +\overset{_\mathrm{(j)}}\Pi_j{}_{\mu\nu}(p)
  +\overset{_\mathrm{(k)}}\Pi_j{}_{\mu\nu}(p)
\right)
\bigg|^{\Lambda^0}
\\=
-\frac{g^2 c_v}{32\pi^2}
\bigg[
   \frac{M^2}{10}C_2\delta_{\mu\nu}
   +\bigg(
      \frac{21}{3}\ln\left(\frac{\pi p}{2M}\right)
     -{7}-\frac{7\pi^2p^2}{72M^2}
    \bigg)p^2\delta_{\mu\nu} \\
   -
    \bigg(
       6\ln\left(\frac{\pi p}{2M}\right)
      -\frac{17}{3}-\frac{4\pi^2 p^2}{45M^2}
    \bigg)p_\mu p_\nu
   +O\left(M^{-4}\right)
\bigg],
\label{eq:summation of (h)(i)(j)(k)}
\end{multline}
and the massless contributions,
\begin{multline}
 -\overset{_\mathrm{(h)}}\Pi_0{}_{\mu\nu}(p)
 -\overset{_\mathrm{(i)}}\Pi_0{}_{\mu\nu}(p)
 -\overset{_\mathrm{(j)}}\Pi_0{}_{\mu\nu}(p)
 -\overset{_\mathrm{(k)}}\Pi_0{}_{\mu\nu}(p)
\bigg|^{\Lambda^0}
\\
=+g^2c_v
\int\frac{\mathrm{d}^4k}{(2\pi)^4}\frac{1}{k^2q^2}
\bigg[
    \left(k^2+q^2-2p^2\right)\delta_{\mu\nu}
    -k_\mu q_\nu - k_\mu k_\nu - q_\mu q_\nu + 2p_\mu p_\nu
\bigg].
\label{eq:massless term of (h)(i)(j)(k)}
\end{multline}
Here we use \eqref{eq:result of infinite sum}
to get the r.h.s.{} of \eqref{eq:summation of (h)(i)(j)(k)}.

    \subsubsection{Ghost field type (C-type) diagrams }
The diagrams containing the ghosts or the PV for the ghosts
are classified into this group.
Since there are some differences in the vertex functions
between the ghost and the PV field,
$\overline c$ and $c$
do not play the role of $\overline c_0$ and $c_0$.
The diagram (e) cannot be identified as the zeroth diagram of (f)
though the both diagrams are the same in graphically.
For the diagram (g), the situation is the same as B-type diagrams.
So we take the same care 
in the similar way as \eqref{eq:cure for no massless diagrams}
to extract the massive contribution.
Then the massive contribution of this group is
\begin{multline}
\sum_{i=-\infty}^\infty
(-1)^i
\left(
   \overset{_\mathrm{(f)}}\Pi_i{}_{\mu\nu}(p)
  +\overset{_\mathrm{(g)}}\Pi_i{}_{\mu\nu}(p)
\right)
\bigg|^{\Lambda^0}
\\=
-\frac{g^2 c_v}{32\pi^2}
   \bigg(
      \frac{4}{3}\ln\left(\frac{\pi p}{2m}\right)
     -\frac{16}{9}-\frac{\pi^2p^2}{90m^2}
   \bigg)
   \left(p^2\delta_{\mu\nu}-p_\mu p_\nu \right)
   +O\left(m^{-4}\right),
\label{eq:summation of (f)(g)}
\end{multline}
and the massless contribution is
\begin{multline}
   \overset{_\mathrm{(e)}}\Pi{}_{\mu\nu}(p)
 - \overset{_\mathrm{(f)}}\Pi_0{}_{\mu\nu}(p)
 - \overset{_\mathrm{(g)}}\Pi_0{}_{\mu\nu}(p)
\bigg|^{\Lambda^0}
\\=
-g^2c_v
\int\frac{\mathrm{d}^4k}{(2\pi)^4}\frac{1}{k^2q^2}
\bigg[
    2q^2\delta_{\mu\nu}
    - k_\mu k_\nu - q_\mu q_\nu - q_\mu k_\nu
\bigg].
\label{eq:massless term of (f)(g)}
\end{multline}
Both the massive contributions from (f) and (g) give
the quadratic divergence proportional to $C_2$
after the use of \eqref{eq:result of infinite sum}.
These contributions are canceled out
because they have the same value except the sign,
and then the quadratic divergence does not appear
in \eqref{eq:summation of (f)(g)}.

    \subsubsection{Total contribution}
%
We can get the total contribution of the vacuum-polarization tensor
adding all the contributions calculated above.
All the quadratic divergence is canceled out exactly
and then only the logarithmic divergence remains as follows:
\begin{multline}
\overset{_\mathrm{Total}}\Pi_{\mu\nu}(p)
\bigg|^{\Lambda^0}
=
-\frac{g^2c_v}{8\pi^2}
  \bigg(
    \frac{10}{3}\ln\left(\frac{\pi p}{2M}\right)
   -\frac{28}{9}-\frac{\pi^2 p^2}{20M^2}
  \bigg)
  \left(p^2\delta_{\mu\nu}-p_\mu p_\nu \right)
  +O\left(M^{-4}\right)
\\
-\frac{g^2c_v}{8\pi^2}
  \bigg(
    \frac{1}{3}\ln\left(\frac{\pi p}{2m}\right)
   -\frac{4}{9}-\frac{\pi^2 p^2}{360m^2}
  \bigg)
  \left(p^2\delta_{\mu\nu}-p_\mu p_\nu \right)
  +O\left(m^{-4}\right)
\\
  +\frac{g^2c_v}{8\pi^2}
     \left(p^2\delta_{\mu\nu}-p_\mu p_\nu \right)
     \left(\ln p^2 + \mathcal{C}_1\right),
\label{eq:total contribution at naively limit of Lambda}
\end{multline}
where the first line comes from the infinite sum
\eqref{eq:summation of (a) and (b)}, \eqref{eq:summation of (c) and (d)} and 
\eqref{eq:summation of (h)(i)(j)(k)} and
the second line from the infinite sum \eqref{eq:summation of (f)(g)}.
We use \eqref{eq:N=1,D=4,x=0,l=-1}
to derive the third line which is
from the massless term \eqref{eq:massless term of (h)(i)(j)(k)}
and \eqref{eq:massless term of (f)(g)}.
As we see in \eqref{eq:estimation of constants}
it is estimated to have a logarithmic divergence
such as $\mathcal{C}_1=-\ln \mathcal{M}^2$
where $\mathcal{M}$ is a parameter of the dimension of one.
If we choose $M=m=\mathcal{M}$,
the divergent part of the vacuum polarization tensor is calculated
\begin{equation}
\overset{_\mathrm{Total}}\Pi_{\mu\nu}(p)\bigg|_\mathrm{div}^{\Lambda^0}
=
\frac{g^2c_v}{8\pi^2}
 \frac{5}{3}
\ln M
(p^2\delta_{\mu\nu} - p_\mu p_\nu).
\label{eq:divergent part of VPT at naively limit of Lambda}
\end{equation}
This result corresponds to the usual logarithmic divergence of the YM theory
at the Feynman gauge.

%
Here we discuss how to cancel the quadratic divergence in the above calculation
analyzing the contribution from each group of diagrams.
We have divided the diagrams into the three types, A, B and C.
The contribution from the massless part does not appear in A-type
because the gauge field is identified as the zeroth PV field
and all the contribution from this type is calculated
as the massive contribution.
This situation is the same as the matter field in the chiral gauge theory.
On the other hand,
$b$, $c$ and $\overline c$ field is not identified
as $b_0$, $c_0$ and $\overline c_0$ respectively,
not only the massive contribution but the massless one arises
in B- and C-type.
In C-type, however, the quadratic divergence from the massive contribution
is cancelled after the summation of the infinite diagrams
and then the quadratic divergence only comes from the massless one.
All the arising aspects of the quadratic divergence are listed
in Table~\ref{tab:quadratic divergence} of Section~4.

The cancellation of the massive contribution occurs
between A- and B-type.
Notice that
B-type will be included in A-type
after $b$ and $b_j$ are integrated out from the theory,
we see this cancellation is essentially the same 
as the cancellation of massive contribution in C-type:
the cancellation occurs in the same group
and the quadratic divergence does not appear in outside.

On the other hand,
the massless contribution is cancelled between B- and C-type.
We remember that
all the massless contribution essentially comes from
lacking terms to take the infinite sum of the PV diagram,
this cancellation means that
the lacking term does not give the quadratic divergence in total
although each diagram does.

  \subsection{Ghost self-energy and vertex correction}
In the above subsection,
we show that the vacuum polarization tensor is regularized
except the logarithmic divergence.
Our next task is to treat this divergence with the renormalization procedure.
The simplest way to renormalize the YM theory
is to calculate the contributions of the ghost self-energy
$\Omega(p)$ and of the gauge-ghost-ghost vertex
$(\Gamma_{Ac\overline c})_\mu(p)$.
In this subsection,
we calculate these contributions at one-loop level
to give a renormalization in the following.

\begin{figure}[htbp]
\begin{center}
\includegraphics
{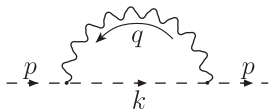}
\caption
{The ghost self energy loop}
\label{fig:4D ghost self energy}
\end{center}
\end{figure}
We cite the diagram of the ghost self-energy
in Figure~\ref{fig:4D ghost self energy}.
Since there is no contribution from the PV fields
we can easily calculate the diagram using
\eqref{eq:N=1,D=4,x=0,l=-1} as follows:
\begin{equation}
\Omega(p) 
\bigg|_\mathrm{div}^{\Lambda^0}
=
-g^2c_v
\int\frac{\mathrm{d}^4k}{(2\pi)^4}
\frac{kp}{k^2q^2}
=-\frac{g^2c_v}{16\pi^2}p^2\ln\mathcal{M}.
\label{eq:divergence from ghost-self energy}
\end{equation}
In the same way,
the one-loop corrections to the gauge-ghost-ghost vertex
in Figure~\ref{fig:4D ghost-ghost-gauge}
are calculated as follows:
\begin{figure}[htbp]
\begin{center}
\includegraphics
{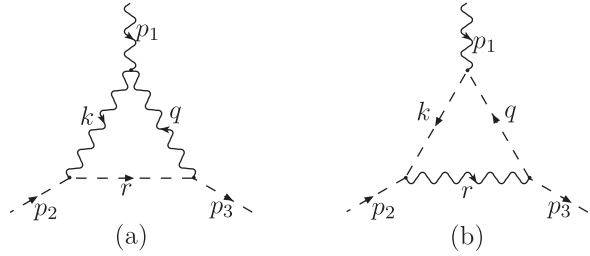}
\caption
{The one loop correction to $(\Gamma_{Ac\overline c})_\mu$.}
\label{fig:4D ghost-ghost-gauge}
\end{center}
\end{figure}
\begin{equation}
(\Gamma_{Ac\overline c})_\mu(p)
\bigg|_\mathrm{div}^{\Lambda^0}
=
-\frac{\mathrm{i}g^3c_v}{32\pi^2}p_\mu
\ln \mathcal{M}^2.
\label{eq:divergent part of g-g-g}
\end{equation}
Here we use \eqref{eq:N=1,D=4,x=0,l=-1} again.

  \subsection{Renormalization}
\label{sec:renormalization of YM}

Now we give a renormalization procedure
to absorb the logarithmic divergence
which we calculate in the preceding subsections.
Since we are only considering the one-loop corrections,
the usual renormalization procedure of the YM theory can be used here.
In that procedure, it is well known that
all the divergences are renormalized
by the three renormalization constants $z_1$, $z_3$ and $z_c$
as follows:
\begin{equation}
\begin{aligned}
A_\mu{}_\mathrm{bare} &= z_3^{\frac{1}{2}}A_\mu,\qquad&
c_\mathrm{bare} &= z_c^{\frac{1}{2}}c,\qquad&
\overline c_\mathrm{bare} &= z_c^{\frac{1}{2}}\overline c,\\
b_\mathrm{bare} &= z_3^{-\frac 1 2}b,\qquad&
g_\mathrm{bare}&=z_1 z_3^{-\frac 3 2}g,\qquad&
\xi_0{}_\mathrm{bare} &= z_3 \xi_0.&
\end{aligned}
\label{eq:renormalization}
\end{equation}
Here we denote the bare parameters with the index `bare'.
Expanding the renormalization constants with $\hbar$ such as
$z_i=\sum_n \hbar^n z_i^{(n)}=1+\hbar z_i^{(1)} + O(\hbar^2)$,
the first order of $\hbar$ corresponds to the one-loop corrections.
Then we get the following equations:
\begin{subequations}
\begin{align}
\Pi_{\mu\nu}(p)\Big|_\mathrm{div}^{\Lambda^0}&=
z_3^{(1)}(p^2\delta_{\mu\nu}-p_\mu p_\nu), \\
\Omega(p)\Big|_\mathrm{div}^{\Lambda^0}&=
-z_c^{(1)}p^2,\\
(\Gamma_{Ac\overline c})_\mu (p)\Big|_\mathrm{div}^{\Lambda^0}
&=
\mathrm{i}g(z_1^{(1)}+z_c^{(1)}-z_3^{(1)})p_\mu.
\end{align}
\label{eq:equation for renormalization factor}
\end{subequations}
Comparing these equations with the results
\eqref{eq:divergent part of VPT at naively limit of Lambda},
\eqref{eq:divergence from ghost-self energy} and
\eqref{eq:divergent part of g-g-g},
the renormalization constants are easily calculated as follows:
%
\begin{align}
z_1^{(1)}\big|_{\xi_0=1}
&=\frac{g^2 c_v}{16\pi^2}\frac{4}{3}  \ln M,&
z_3^{(1)}\big|_{\xi_0=1}
&=\frac{g^2 c_v}{16\pi^2}\frac{10}{3} \ln M,&
z_c^{(1)}\big|_{\xi_0=1}
&=\frac{g^2 c_v}{16\pi^2}\ln M.
\label{eq:renormalization constants}
\end{align}
From these results,
the $\beta$- and $\gamma$-function at a renormalization point $\mu$
are calculated
\begin{subequations}
\begin{align}
\beta (g,\xi_0)\big|_{\xi_0=1}
&= \mu \frac{\partial g}{\partial \mu}
= - \frac{g^3}{16\pi^2}\frac{11}{3}c_v + O(g^5), \\
\gamma_A (g,\xi_0)\big|_{\xi_0=1}
&= \frac{\mu}{2}\frac{\partial \ln z_3}{\partial \mu}
= - \frac{g^2}{16\pi^2}\frac{5}{3}c_v + O(g^4).
\end{align}
\end{subequations}
These values accord with the familiar value at Feynman gauge%
~\cite{Politzer,Gross-Wilczek73PRL,Gross-Wilczek73PhysRev}.

\section{Contributions from HCD Terms}
\label{sec:one-loop contributions in finite}

In the last section,
we have seen that
the quadratic divergence is completely canceled
and the correct RG $\beta$- and $\gamma$-function are given
with our regularization scheme
by the calculation of the one-loop contributions
in $\Lambda^0$ order.
In this section,
we mainly consider the contribution from $\Lambda$-dependent part.

We remind that
the HCD terms render the renormalizable theory into a
super-renormalizable one,
it is reasonable to consider that
all the divergence from higher than two-loop level
in the renormalizable theory
is translated into some divergence at one-loop level
in the super-renormalizable one.
So there is some divergence in $\Lambda$-dependent terms
and it must be regularized by our regulators.
We now check such divergence is regularized at least in $\Lambda^{-4}$ order.

  \subsection{Vacuum polarization tensor in $\Lambda^{-4}$ order}

\begin{figure}[htbp]
\begin{center}
\includegraphics
{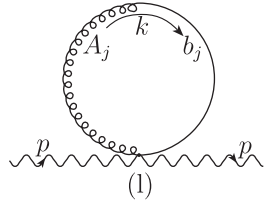}
\caption
{Diagram appears in the $\Lambda^{-4}$ order.}
\label{fig:loops only Lambda4}
\end{center}
\end{figure}

A new diagram listed in Figure~\ref{fig:loops only Lambda4}
arises in this order
in addition to the diagrams in Figure~\ref{fig:loops}.
This is from the irrelevant vertex
which maintains the BRST invariance.
We classify this diagram into B-type
because it contains $b_j$ field in the internal line.
The contribution is calculated in a similar way
with the case of $\Lambda^0$ order
under the rules mentioned
on the top of Section~\ref{sec:one-loop contributions}.
Since the contribution is divided into two parts
following the calculation rules,
one is the infinite sums and the other is the massless terms,
all our tasks are concentrated on the calculations of the infinite sums
(arising from the PV fields like \eqref{eq:summation of (h)(i)(j)(k)})
and of the massless terms
(from the counter terms such as \eqref{eq:massless term of (h)(i)(j)(k)}).

For the contribution of the infinite sums,
all the terms are written by the following formula:
\begin{multline}
\sum_{j=-\infty}^\infty
(-1)^j
\frac{M^{2r}_j}{\Lambda^4}
\int \frac{\mathrm{d}^4k}{(2\pi)^4}
\frac{
      k_{\mu_1}\cdots k_{\mu_{10-2r-s}}
      p_{\mu_{10-2r-s+1}}\cdots p_{\mu_{10-2r}}
      }
      {(k^2+M_j^2)^2(q^2+M_j^2)^2}
\\
\sim
\frac{1}{\Lambda^4}
\left(
 M^{2\left[3-\frac{s}{2}\right]}
 +O\left(M^{2\left[2-\frac{s}{2}\right]}\right)
\right),
\label{eq:infinite sum at lambda4}
\end{multline}
where $r$ and $s$ denote the order of $M^2$ and $p_\mu$
in the integrand,
and run in the region $0\le r\le 5$
and $0\le s\le 10-2r$ respectively.
We use the computation rules discussed
in Appendix~\ref{appendix:calculations of infinite sum}
to get the second line
and $[\quad]$ denotes the Gauss' notation.
Its contribution depends on the order of the limit $M$ and $\Lambda$,
but the difference does not affect renormalization group functions%
~\cite{Martin-Ruiz95},
we may restrict our calculations within the ratio
${M}/{\Lambda}=\mathrm{constant}$.
Then \eqref{eq:infinite sum at lambda4} behaves
\begin{equation}
\sim
M^{2\left[1-\frac{s}{2}\right]}
+O\left(M^{2\left[-\frac{s}{2}\right]}\right),
\end{equation}
and only the terms having $s=0$ give the divergence
in the limit $M\rightarrow\infty$,
they are calculated in \eqref{eq:result of infinite sum N=2}.

Similarly the contributions from the massless terms
are written in the following form
using the formula \eqref{eq:N=1,D=4,x=0,general}
and \eqref{eq:estimation of constants},
\begin{multline}
\frac{1}{\Lambda^4}
\int\frac{\mathrm{d}^4k}{(2\pi)^4}
\frac{k_{\mu_1}\cdots k_{\mu_{6-s}}p_{\mu_{6-s+1}}\cdots p_{\mu_6}}
     {k^2q^2}
\\
\sim
\frac{1}{\Lambda^4}
\left(
  \mathcal{M}^{2\left[3-\frac{s}{2}\right]}+
  p^2\mathcal{M}^{2\left[2-\frac{s}{2}\right]}+
  \cdots+
  p^{2\left[3-\frac{s}{2}\right]}\ln \mathcal{M}^2+
  \cdots
\right)
\label{eq:estimation of counter term in lambda4}
\end{multline}
where $0\le s \le 6$.
Also in this case, only $s=0$ terms give the divergence
under the condition ${M}/{\Lambda}=\mathrm{constant}$
if we identify $\mathcal{M}$ with $M$.

Both in \eqref{eq:infinite sum at lambda4} and
\eqref{eq:estimation of counter term in lambda4},
the logarithmic divergence appears in the form of $\ln M^2 /\Lambda^4$.
Since it vanishes in the ratio $M/\Lambda = \mathrm{constant}$
there is no logarithmic divergence
and only the terms containing no external momentum $p_\mu$
in the numerator give the quadratic divergence
in this order.
So we only consider $s=0$ terms
and confirm whether the divergence cancels or not
by calculating each diagrams.

Now we calculate the divergent contributions in the below.
For the diagrams in A-type,
since the diagram (a) and (c) play the massless diagram respectively
we take the infinite sum without any extra diagram
as in the $\Lambda^0$ order.
The divergent contributions from (a), (b), (c) and (d)  are
\begin{equation}
\sum_{j=-\infty}^\infty
(-1)^j
\left(
   \overset{_\mathrm{(b)}}\Pi_j{}_{\mu\nu}(p)
  +\overset{_\mathrm{(d)}}\Pi_j{}_{\mu\nu}(p)
\right)\bigg|_\mathrm{div}^{\Lambda^{-4}} =
-\frac{g^2 c_v }{8\pi^2}
\frac{9}{154}M^2C_4
\delta_{\mu\nu}.
\label{eq:quadratic divergence related to the gauge}
\end{equation}
Here we use \eqref{eq:result of infinite sum N=2}
and $C_4$ is a dimensionless constant.

For the diagrams in C- and B-type,
we take the infinite sum adding the external diagrams
as in \eqref{eq:cure for no massless diagrams}
\begin{multline}
\overset{_\mathrm{(e)}}\Pi{}_{\mu\nu}(p)
+
\sum_{\substack{i=-\infty \\ i\ne 0}}^\infty
(-1)^i
\left(
   \overset{_\mathrm{(f)}}\Pi_i{}_{\mu\nu}(p)
  +\overset{_\mathrm{(g)}}\Pi_i{}_{\mu\nu}(p)
\right)\bigg|_\mathrm{div}^{\Lambda^{-4}}
\\
=
-\frac{g^2c_v}{\Lambda^4}
 \int \frac{\mathrm{d}^4 k}{(2\pi)^4}\frac{1}{k^2q^2}
      \left( 2k^6\delta_{\mu\nu} - 4k^4k_\mu k_\nu \right),
\label{eq:quadratic divergence related to the ghosts}
\end{multline}
\begin{multline}
\sum_{\substack{j=-\infty \\ j\ne 0}}^\infty
(-1)^j
\left(
   \overset{_\mathrm{(h)}}\Pi_j{}_{\mu\nu}(p)
  +\overset{_\mathrm{(i)}}\Pi_j{}_{\mu\nu}(p)
  +\overset{_\mathrm{(j)}}\Pi_j{}_{\mu\nu}(p)
  +\overset{_\mathrm{(k)}}\Pi_j{}_{\mu\nu}(p)
  +\overset{_\mathrm{(l)}}\Pi_j{}_{\mu\nu}(p)
\right)\bigg|_\mathrm{div}^{\Lambda^{-4}}
\\
=
\frac{g^2 c_v }{8\pi^2}
\frac{9}{154}M^2C_4\delta_{\mu\nu}
+\frac{g^2c_v}{\Lambda^4}
 \int \frac{\mathrm{d}^4 k}{(2\pi)^4}\frac{1}{k^2q^2}
      \left( 2k^6\delta_{\mu\nu} - 4k^4k_\mu k_\nu \right).
\label{eq:quadratic divergence related to b_j}
\end{multline}
The first term of the r.h.s.{}
in \eqref{eq:quadratic divergence related to b_j}
arises from the infinite sums of PV diagrams,
and the second term from the massless terms.
In \eqref{eq:quadratic divergence related to the ghosts},
we add the contribution from the diagram (e).
The each diagram (f) and (g) gives the quadratic divergence
proportional to the constant $C_4$
with the usage of \eqref{eq:result of infinite sum N=2},
but they have the opposite sign and cancel each other.
Then only the quadratic divergence from the massless terms remains.
This situation is the same as we see in \eqref{eq:massless term of (f)(g)}
and also at the $\Lambda^0$ order.

All the quadratic divergence of the order of $\Lambda^0$ and $\Lambda^{-4}$
are listed in Table~\ref{tab:quadratic divergence}.
\begin{table}[tbp]
\def\arraystretch{1.5}
\begin{center}
\begin{tabular}{ll|cc|cc}
\noalign{\hrule height 0.8pt}
           \multicolumn{2}{c|}
           {\smash{%
             \raisebox{-.5\normalbaselineskip}{Diagram type}
            }}
          &\multicolumn{2}{c|}{$\Lambda^0$ order}
          &\multicolumn{2}{c}{$\Lambda^{-4}$ order}
\\ \cline{3-6}
           \multicolumn{2}{c|}{}
          &{massive}
          &{massless}
          &{massive}
          &{massless}
\\ \noalign{\hrule height 0.8pt}
           \smash{%
             \raisebox{-.5\normalbaselineskip}{Gauge}
           }
          & A-type
          & $+\frac{1}{10}M^2 C_2$
          & $0$
          & ${-\frac{18}{77\Lambda^4}M^6 C_4}$
          & $0$
          \\
          & B-type
          & $-\frac{1}{10}M^2 C_2$
          & $+\frac{5}{6}\mathcal{M}^2$
          & ${+\frac{18}{77\Lambda^4}M^2 C_4}$
          & ${+\frac{2}{105\Lambda^4}\mathcal{M}^6}$
          \\ \hline
            Ghost
          & C-type
          & $0$
          & $-\frac{5}{6}\mathcal{M}^2$
          & $0$
          & ${-\frac{2}{105\Lambda^4}\mathcal{M}^6}$
          \\
\noalign{\hrule height 0.8pt} 
\end{tabular}
\caption
{%
 Quadratic divergence at $\Lambda^0$ and $\Lambda^{-4}$ order.
 `massive' and `massless' means the contribution from
 the infinite sum such as \eqref{eq:summation of (h)(i)(j)(k)} and
 the massless terms such as \eqref{eq:massless term of (h)(i)(j)(k)}
 respectively.
 The coefficient $\frac{g^2c_v}{32\pi^2}\delta_{\mu\nu}$ are abbreviated
 from explicit values.
 }
\label{tab:quadratic divergence}
\end{center}
\end{table}
It is easy to see from
\eqref{eq:quadratic divergence related to the gauge},
\eqref{eq:quadratic divergence related to the ghosts} and
\eqref{eq:quadratic divergence related to b_j}
that all the divergence of the vacuum polarization tensor
at $\Lambda^{-4}$ order disappears from the theory.
It clearly shows that the divergence at this order is removed completely.
This cancellation mechanism is exactly the same
as we observed in Section~\ref{sec:one-loop contributions}:
B-type contribution gives the counter terms
for A- and C-type.
Considering A-type diagrams do not give any massless contribution
in all orders of $\Lambda$,
the same mechanism is strongly expected
to work in the higher order of $\Lambda^{-4}$,
e.g. in $\Lambda^{-8}$ order.

Such a cancellation mechanism is observed
without an explicit identification of the parameters
$M=m=\mathcal{M}$
which is necessary for the renormalization,
because the cancellation is occurred between the divergence
described by the same parameters:
massive divergence is canceled by massive one
and massless divergence by massless one.

  \subsection{Inconsistent higher derivative action}
  \label{sec:Inconsistent higher derivative action}

There is another choice of gauge-fixing action but \eqref{eq:GF with HD action}.
Remembering that the higher derivative function for the gauge-fixing action
is added to improve the convergence of the longitudinal part of
the gauge propagator,
we can choose the action with a higher derivative function $f$
according to the references~\cite{Martin-Ruiz95,Martin-Ruiz95PL}
as follows;
\begin{equation}
S_\mathrm{GF}^{f} =
\int \mathrm{d}^4x\,
 b\frac{\xi_0 }{2f^2}b
 -b \partial^\mu A_\mu
 +\overline c  \partial_\mu D^\mu c.
\label{eq:gauge-fixing action with f}
\end{equation}
This action is extended to the Pauli-Villars field
using a higher \textit{covariant} derivative function $\Tilde f$,
\begin{equation}
S_{b_j}^{f} =
\int \mathrm{d}^4 x
 \left[
  b_j\frac{\xi_j}{2\Tilde f^2}b_j -b_jD^\mu A_j{}_\mu
 \right].
\label{eq:gauge-fixing action with f for PV}
\end{equation}
These actions give the same propagators for the gauge field and PV for gauge
as we get with the actions
\eqref{eq:GF with HD action} and \eqref{eq:auxiliary field for PV action}
under the condition $f=H$ and $\Tilde f = \Tilde H$,
but change some propagators and vertices.
Following points are modified by the usage of the action
\eqref{eq:gauge-fixing action with f} and
\eqref{eq:gauge-fixing action with f for PV};
\begin{enumerate}
\item
The $\Lambda$-dependent parts of
propagators $\left<A_jb_j\right>$ and $\left<b_jb_j\right>$
and vertices $\left<Ab_jA_j\right>$ and $\left<AAb_jA_j\right>$
are modified.
Especially, these vertices do not give any $\Lambda$-dependent term;

\item
New vertices $\left<Ab_jb_j\right>$ and $\left<AAb_jb_j\right>$
appear because of $\Tilde f$.
These vertices give the new diagrams.

\item
The ghost and its PV field have no $\Lambda$-dependence.

\end{enumerate}

From the first fact,
\eqref{eq:quadratic divergence related to b_j},
which is the total contribution of the diagrams (h), (i), (j), (k) and (l),
is changed.
The diagram (l), especially, does not appear
because of the absence of the vertex $\left<AAb_jA_j\right>$. 
Recalculating these diagrams with modified propagators and vertices,
we get
\begin{multline}
\sum_{\substack{j=-\infty \\ j\ne 0}}^\infty
(-1)^j
\left(
   \overset{_\mathrm{(h)}}\Pi_j{}_{\mu\nu}(p)
  +\overset{_\mathrm{(i)}}\Pi_j{}_{\mu\nu}(p)
  +\overset{_\mathrm{(j)}}\Pi_j{}_{\mu\nu}(p)
  +\overset{_\mathrm{(k)}}\Pi_j{}_{\mu\nu}(p)
\right)\bigg|_\mathrm{div}^{\Lambda^{-4}}
\\
=
\frac{g^2 c_v }{8\pi^2}
\frac{6}{154}M^2C_4\delta_{\mu\nu}.
\label{eq:quadratic divergence with f}
\end{multline}

\begin{figure}[tbp]
\begin{center}
\includegraphics
{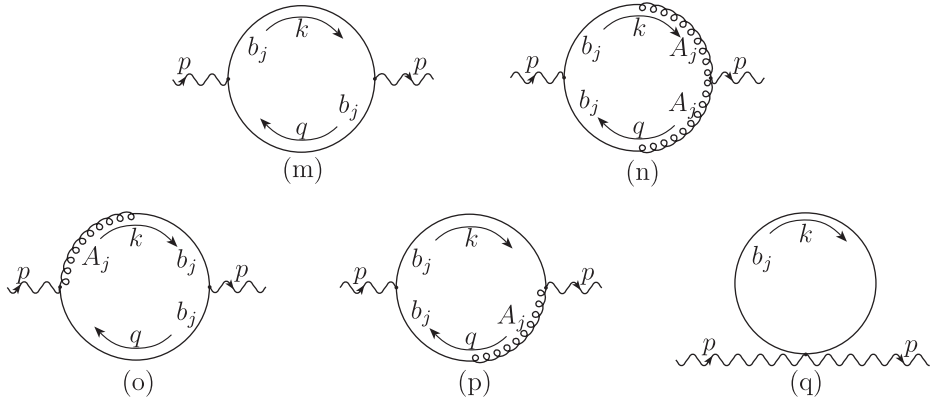}
\caption
{Diagrams generated by the new vertices appearing in
\eqref{eq:gauge-fixing action with f} and
\eqref{eq:gauge-fixing action with f for PV}.
These diagrams only contribute to the $\Lambda$ depending terms
but do not give any quadratic divergence to $\Lambda^{-4}$ in total.
}
\label{fig:alter diagrams}
\end{center}
\end{figure}

The second fact leads the new diagrams listed
in  Figure~\ref{fig:alter diagrams}.
Since the diagram (m) only gives the higher order of $\Lambda^{-8}$
and the diagram (n) does not give any quadratic divergence,
the diagrams we have to calculate here are the three diagrams of
(o), (p) and (q).
Each diagram gives the quadratic divergence after the infinite sum
but all of them are canceled in total,
the massive contribution does not appear from these new diagrams.
Not only the massive contribution but massless one is not arising
from these diagrams:
all these diagrams contain mass parameter $M_j$ in the numerator of integrand
so their zeroth diagrams vanish taking the massless limit of them,
like the diagram (k) in Section~\ref{sec:one-loop contributions}.

Since the ghost and its PV do not give
any $\Lambda$-dependent contribution in this case,
the quadratic divergence in the $\Lambda^{-4}$ order
only comes from
\eqref{eq:quadratic divergence related to the gauge}
and \eqref{eq:quadratic divergence with f}.
Comparing these two contributions,
we find the quadratic divergence does not cancel
in this order.

%
The reason why
the action \eqref{eq:gauge-fixing action with f} and
\eqref{eq:gauge-fixing action with f for PV}
fail the cancellation of the quadratic divergence in $\Lambda^{-4}$
comes from the fact that
the function $\Tilde f$ gives a different effect to the mass term
of the PV field for ghost,
with the case of $\Tilde H$.
Certainly,
\eqref{eq:gauge-fixing action with f}
(\eqref{eq:gauge-fixing action with f for PV})
is translated into \eqref{eq:GF with HD action}
(\eqref{eq:auxiliary field for PV action})
by the redefinition of $\overline c \rightarrow \overline c H$
($\overline c_i \rightarrow \overline c_i \Tilde H$)
and $b \rightarrow b H$ ($b_j \rightarrow b_j \Tilde H$)
under the condition $f=H$ ($\Tilde f = \Tilde H$),
but the PV for ghost does not:
the mass term gets the $\Lambda$-dependence
and $\det{}^{\gamma_i}\mathbf{C}_i$ does not coincide with \eqref{eq:det C}.
This fact says that
as long as we construct PV fields by simply adding an usual mass term,
\eqref{eq:gauge-fixing action with f} and
\eqref{eq:gauge-fixing action with f for PV}
are not the correct higher derivative regulator.
If we want to use these regulators,
we must add a $\Lambda$-dependence to the mass term of the ghost PV field.

From an another point of view,
it is simply recognized that our actions are natural.
The main purpose to introduce a higher (covariant) derivative function
to the gauge-fixing action is to improve the convergence of the propagator
of the gauge field as we explain
in Section~\ref{sec:Higher covariant derivative method}.
So the function must regularize the gauge field.
The function $f$ ($\Tilde f$) surely improves the convergence of the propagator
by regularizing the gauge-fixing parameter $\xi_0$ ($\xi_j$),
but it is not the regulator for the gauge (PV) field.
On the other hand,
the higher (covariant) derivative function $H$ ($\Tilde H$)
regularizes $A_\mu$ ($A_j{}_\mu$)
by multiplying to the cross term with the auxiliary field $b$ ($b_j$)
and gives the complete cancellation of the quadratic divergence
without any $\Lambda$-dependence in the mass term of the ghost PV field.
%

\section{Conclusion and Discussion}
\label{sec:conclusion}

%
In this paper,
we check the consistency of the hybrid regularization
in the four-dimensional Yang-Mills theory
when we use the regularization scheme consists of
the higher covariant derivative term
and an infinitely many Pauli-Villars fields
that we applied to the three-dimensional Chern-Simons gauge theory.
By an explicit calculation of the vacuum polarization tensor,
we get the correct factors of the renormalization group
$\beta$- and $\gamma$-function.
Furthermore, 
the cancellation of the quadratic divergence is also demonstrated
in our method with an expansion of $\Lambda^{-1}$.
These facts show that
our regularization method is available for a divergent theory
not only for a finite theory.

%
In our calculation, all the diagrams are classified into the three types
and their quantum corrections are calculated separately in each type.
In that calculation, all the contributions are separated into
`massive' and `massless' contributions
along the manipulation rules given in Section~\ref{sec:one-loop contributions}.
As a result,
we can clearly confirm the cancellation of the quadratic divergence:
the `massive' contribution is cancelled between A- and B-type diagrams
and the `massless' between B- and C-type.
This clear cancellation mechanism works identically
in both the order of $\Lambda^0$ and $\Lambda^{-4}$
as we listed in Table~\ref{tab:quadratic divergence}.

%
Such a mechanism is expected to occur in the quadratic divergence
in the higher order of $\Lambda^{-1}$, e.g. $\Lambda^{-8}$,
because the massless contribution of A-type is always absent.
This argument is proved
if it is shown that
C-type diagrams do not give the massive contribution
in any order of $\Lambda^{-1}$.

%
The higher (covariant) derivative functions for gauge-fixing terms, 
$H$ and $\Tilde H$, especially, are important
for the cancellation of $\Lambda$-depending contributions.
The actions \eqref{eq:gauge-fixing action with f} and
\eqref{eq:gauge-fixing action with f for PV} sure give
the cancellation in $\Lambda^0$ order,
but do not in $\Lambda^{-4}$.
The reason is that
the mass term of the ghost PV does not accord with ours
under the redefinition of fields.
In other words, the functions $f$ and $\Tilde f$ are
only the higher (covariant) derivative
for the gauge-fixing parameters
and they do not regularize the gauge and its PV fields completely.
Though the higher (covariant) derivative terms are originally introduced
to improve the convergence of the gauge and PV field,
such functions consequently must be inserted
to regularize the gauge and PV field
like the action \eqref{eq:GF with HD action}
and \eqref{eq:auxiliary field for PV action},
as far as we introduce an usual mass term like $m_i^2 \overline c_i c_i$.

In usual case,
some extra regularization so-called `pre-regulator'
is introduced to compare all the divergence,
in our method, however,
such a regularization is not needed.
This is because that our method contains
a scheme corresponding to a pre-regulator.
In our calculation,
all the divergence appears in the constants of integration
which is derived from the differential equations of $X$
like \eqref{eq:differential equation} or
\eqref{eq:rho integration and summation with r=0}.
As a result,
`massive' divergence is characterized by $M$ or $m$
and `massless' by $\mathcal{M}$,
and then we get the correct renormalization
supposing $M=m=\mathcal{M}$.
This assumption will corresponds to the conventional pre-regulator
in reference~\cite{Bakeyev-Slavnov96}
where it is only needed to give a rigorous arguments.
So our method is an alternative procedure
which does not need any pre-regularization.

The very reason that we want to avoid the pre-regulator is
that the procedure may break the invariance
which we would like to preserve in the theory.
But in our method,
since a scheme corresponding to a pre-regularization is
in the usual regularization method,
we do not worry about the symmetry breaking.

%
Finally we comment on the problem of the overlapping divergences;
how to treat the one-loop divergence with external PV fields%
~\cite{Bakeyev-Slavnov96, Warr87_1}.
It is difficult
to remove the divergence with our present method,
however, the PV pair such as \eqref{eq:PV determinant}
will give the key to the problem.
Since both the fields give the same diagram with each external field,
if we can find a PV pair whose diagrams cancel each other
like the parity-odd contributions in reference~\cite{Nittoh-Ebihara98MPLA}
where the pair is connected by the parity-transformation,
all the one-loop divergence with external PV fields are canceled
and the problem will be overcome.
Unfortunately, such a PV pair is not found up to the present.

\bigskip


\section*{Acknowledgments}

I would like to thank Dr.~T.~Ebihara
for many useful discussions on the subject of this paper.

I suddenly received the sad news of Mr.~T.~Sakuma's early grave.
His M.Sc.\ thesis entitled
``Problems of regularization in non-Abelian gauge theory
and a possibility of a new regularization''
motivated me to do this work.
I express my deep regret at his untimely death.

\bigskip


\appendix

\section{Feynman Rules}
\label{appendix:Feynman Rules}
%
  \subsection{Propagators}

    \subsubsection{Propagators for original fields}

\begin{align}
A_\mu^a
\begin{aligned}
\includegraphics{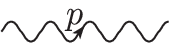}
\end{aligned}
A_\nu^b
&=
\delta^{ab}
\left[
 \frac{\Lambda^4 }{ p^4(p^4+\Lambda^4)}
 (p^2 g_{\mu\nu} - p_\mu p_\nu)
 +\xi_0
 \frac{p_\mu p_\nu }{ p^4 H^2}
\right]
\nonumber \\
&\approx
\frac{\delta^{ab}}{p^2}
\left(
 g_{\mu\nu}+(\xi_0-1)\frac{p_\mu p_\nu}{p^2}
\right)
\left(1-\frac{p^4}{\Lambda^4}\right),
\label{propagator:4D AA}
\end{align}
%
%
\begin{equation}
A_\mu^a
\begin{aligned}
\includegraphics{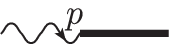}
\end{aligned}
b^b
=
\frac{i\delta^{ab} p_\mu }{ p^2 H}
\approx
\frac{i\delta^{ab} p_\mu }{ p^2}
\left(1-\frac{p^4}{2\Lambda^4}\right),
\label{propagator:4D bA}
\end{equation}
\begin{equation}
c^a
\begin{aligned}
\includegraphics{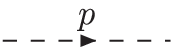}
\end{aligned}
\overline c^b
=
\frac{-\delta^{ab}}{ p^2 H}
\approx
\frac{-\delta^{ab}}{ p^2}
\left(1-\frac{p^4}{2\Lambda^4}\right),
\label{propagator:4D cc}
\end{equation}

    \subsubsection{Propagators for PV fields}

\begin{multline}
A_j{}_\mu^a
\begin{aligned}
\includegraphics{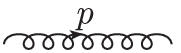}
\end{aligned}
A_j{}_\nu^b
=
\delta^{ab}
\left[
\frac{\Lambda^4 (p^2 g^{\mu\nu} - p^\mu p^\nu)}
{ (\Lambda^4 + p^4)p^4 + M_j^2\Lambda^4p^2}
+
\xi_j \frac{p^\mu p^\nu }
{ p^2(p^2 H^2 + \xi_j M_j^2)}
\right]
\\
\shoveleft{
 \phantom{
  A_j{}_\mu^a
  \begin{aligned}
  \includegraphics{AjAj.eps}
  \end{aligned}
  A_j{}_\nu^b
 }
\approx
\frac{\delta^{ab}}{p^2}
\bigg[
  \frac{p^2g_{\mu\nu}-p_\mu p_\nu}{p^2+M_j^2}
  \left(1-\frac{1}{\Lambda^4}\frac{p^6}{p^2+M_j^2}\right)
}
\\
  +
  \xi_j\frac{p_\mu p_\nu}{p^2+\xi_j M_j^2}
  \left(1-\frac{1}{\Lambda^4}\frac{p^6}{p^2+\xi_jM_j^2}\right)
\bigg],
\label{propagator:4D AjAj}
\end{multline}
\begin{equation}
A_j{}_\mu^a
\begin{aligned}
\includegraphics{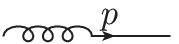}
\end{aligned}
b_j^b
=
\frac{\mathrm{i}\delta^{ab} p_\mu H}{p^2 H^2 + \xi_j M_j^2}
\approx
\frac{\mathrm{i}\delta^{ab} p_\mu}{p^2 + \xi_j M_j^2}
\left(
  1-\frac{p^4}{2\Lambda^4}\frac{p^2-\xi_j M_j^2}{p^2+\xi_j M_j^2}
\right),
\label{propagator:4D bjAj}
\end{equation}
\begin{equation}
b_j^a
\begin{aligned}
\includegraphics{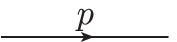}
\end{aligned}
b_j^b
=
\frac{-\delta^{ab}M_j^2 }{p^2 H^2 + \xi_j M_j^2}
\approx
\frac{-\delta^{ab}M_j^2 }{p^2+ \xi_j M_j^2}
\left(
  1-\frac{1}{\Lambda^4}\frac{p^6}{p^2+\xi_j M_j^2}
\right)
,
\label{propagator:4D bjbj}
\end{equation}
\begin{equation}
c_i^a
\begin{aligned}
\includegraphics{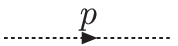}
\end{aligned}
\overline c_i^b
=
\frac{-\delta^{ab}}{ p^2 H + m_i^2}
\approx
\frac{-\delta^{ab}}{ p^2 + m_i^2}
\left(
  1-\frac{1}{2\Lambda^4}\frac{p^6}{p^2+m_i^2}
\right)
,
\label{propagator:4D cici}
\end{equation}

  \subsection{Vertices}

    \subsubsection{Three-point vertices}

\begin{multline}
\begin{aligned}
\includegraphics{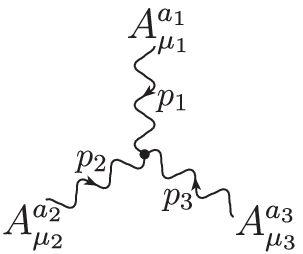}
\end{aligned}
=
\begin{aligned}
\includegraphics{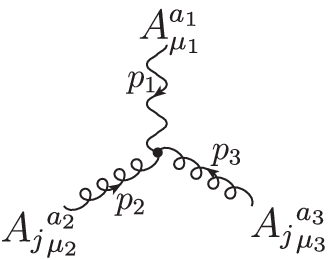}
\end{aligned}
\\
\shoveleft{
=
\frac{\mathrm{i}g}{ \Lambda^4}
f^{a_1a_2a_3}
\Big[
  -\Lambda^4 p_1{}_{\mu_2} g_{\mu_3\mu_1}
  -p_1^4 p_1{}_{\mu_2}g_{\mu_3\mu_1}
  +p_1^2 (p_3-p_1)_{\mu_2}
   (p_1{}_{\mu_3}p_3{}_{\mu_1}-p_1p_3g_{\mu_3\mu_1})
\Big]_\mathrm{sym}
}
\\
\shoveleft{
=
\frac{\mathrm{i}g}{ \Lambda^4}f^{a_1a_2a_3}
\Big[
 -\Lambda^4
 \left\{
   (p_1-p_3)_{\mu_2}g_{\mu_3\mu_1}
  +(p_3-p_2)_{\mu_1}g_{\mu_2\mu_3}
  +(p_2-p_1)_{\mu_3}g_{\mu_1\mu_2}
 \right\}
}
\\
 -(p_1^4p_1-p_3^4p_3)_{\mu_2}g_{\mu_3\mu_1}
 -(p_3^4p_3-p_2^4p_2)_{\mu_1}g_{\mu_2\mu_3}
 -(p_2^4p_2-p_1^4p_1)_{\mu_3}g_{\mu_1\mu_2}
\\
 +(p_1^2+p_3^2)(p_3-p_1)_{\mu_2}
  (p_3{}_{\mu_1}p_1{}_{\mu_3}-p_3p_1g_{\mu_1\mu_3})
\\
 +(p_3^2+p_2^2)(p_2-p_3)_{\mu_1}
  (p_2{}_{\mu_3}p_3{}_{\mu_2}-p_2p_3g_{\mu_3\mu_2})
\\
 +(p_2^2+p_1^2)(p_1-p_2)_{\mu_3}
  (p_1{}_{\mu_2}p_2{}_{\mu_1}-p_1p_2g_{\mu_2\mu_1})
\Big],
\label{vertex:4D AAA}
\end{multline}
where $[\quad]_\mathrm{sym}$ implies that symmetrize all the indices.

\begin{multline}
\begin{aligned}
\includegraphics{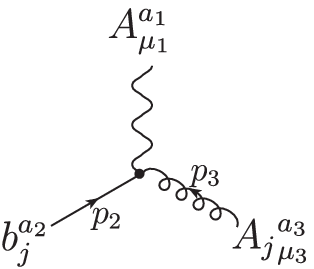}
\end{aligned}
\\
=
gf^{a_1a_2a_3}
\bigg[
  g_{\mu_1\mu_3}
  +
  \frac{1}{2\Lambda^4}
    \left(
      p_2^4g_{\mu_1\mu_3}+(p_2^2+p_3^2)(p_3-p_2)_{\mu_1}p_3{}_{\mu_3}
    \right)
\bigg]
\label{vertex:4D AbjAj}
\end{multline}

\begin{equation}
\begin{aligned}
\includegraphics{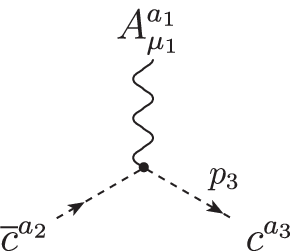}
\end{aligned}
=
-\mathrm{i}gf^{a_1a_2a_3}p_3{}_{\mu_1}
\left( 1+\frac{p_3^4}{2\Lambda^4} \right)
\label{vertex:4D Acc}
\end{equation}

\begin{equation}
\begin{aligned}
\includegraphics{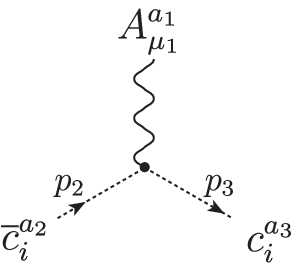}
\end{aligned}
=
-\mathrm{i}gf^{a_1a_2a_3}(p_2+p_3)_{\mu_1}
\left[ 1+ \frac{1}{2\Lambda^4}(p_3^4+p_3^2p_2^2+p_2^4) \right]
\label{vertex:4D Acjcj},
\end{equation}

    \subsubsection{Four-point vertices}

\begin{multline}
\begin{aligned}
\includegraphics{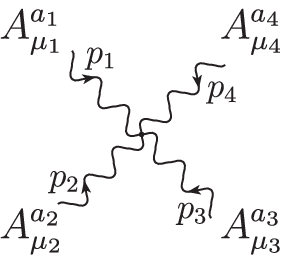}
\end{aligned}
=
\begin{aligned}
\includegraphics{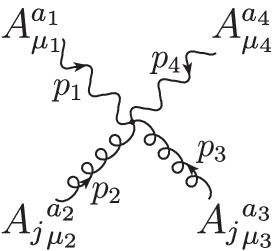}
\end{aligned}
\\
\shoveleft{
=
\frac{-g^2}{ \Lambda^4}
f^{a_1a_2b}f^{a_3a_4b}
\Big[
 \Lambda^4
 (g_{\mu_1\mu_3}g_{\mu_2\mu_4}
  -g_{\mu_1\mu_4}g_{\mu_2\mu_3})
 +
 (p_1+p_2)^4
 (g_{\mu_1\mu_3}g_{\mu_2\mu_4}
  -g_{\mu_1\mu_4}g_{\mu_2\mu_3})
}\\
+8(p_1+p_2)_{\mu_1}
 \left\{
  p_4{}_{\mu_3}
  (p_2p_4g_{\mu_2\mu_4}-p_2{}_{\mu_4}p_4{}_{\mu_2})
  -
  p_3{}_{\mu_4}
  (p_2p_3g_{\mu_2\mu_3}-p_2{}_{\mu_3}p_3{}_{\mu_2})
 \right\}\\
+4p_1^2
 \left\{
  g_{\mu_2\mu_3}
  (p_4{}_{\mu_1}p_1{}_{\mu_4}+p_1p_4g_{\mu_1\mu_4})
  -
  g_{\mu_2\mu_4}
  (p_3{}_{\mu_1}p_1{}_{\mu_3}+p_1p_3g_{\mu_1\mu_3})
 \right\}
\\
 -4p_1^2(2p_1+p_2)_{\mu_2}
 (g_{\mu_1\mu_4}p_1{}_{\mu_3}-g_{\mu_1\mu_3}p_1{}_{\mu_4})
\\
+2p_1{}_{\mu_1}
 \left\{
  p_3{}_{\mu_3}
  (p_2p_4g_{\mu_2\mu_4}-p_2{}_{\mu_4}p_4{}_{\mu_2})
  -
  p_4{}_{\mu_4}
  (p_2p_3g_{\mu_2\mu_3}-p_2{}_{\mu_3}p_3{}_{\mu_2})
 \right\}
\\
+4(p_1+p_2)^2
 \left\{
  g_{\mu_2\mu_4}p_4{}_{\mu_1}(p_3+2p_4)_{\mu_3}
  -
  g_{\mu_2\mu_3}p_3{}_{\mu_1}(p_4+2p_3)_{\mu_4}
 \right\}
\Big]_\mathrm{sym},
\label{vertex:4D AAAA}
\end{multline}

\begin{multline}
\includegraphics{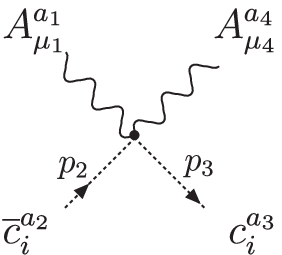}\\
\shoveleft{
=
g^2f^{a_1a_2b}f^{a_4a_3b}
\bigg[
g_{\mu_1\mu_4}
+
\frac{1}{2\Lambda^4}
  \bigg\{
    (p_2^4+p_2^2 p_3^2 +p_3^2)g_{\mu_1\mu_4}
}\\
    +
    (p_2^2 +(p_1+p_2)^2 +p_3^2)
    (p_2+p_3-p_4)_{\mu_1}(p_2+p_3+p_1)_{\mu_4}
  \bigg\}
\bigg]
+
\left(1\leftrightarrow 4\right)
\label{vertex:4D AAcc}
\end{multline}

\begin{multline}
\includegraphics{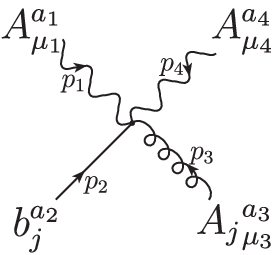}\\
\shoveleft{
=
\frac{\mathrm{i}g^2}{2\Lambda^4}
f^{a_1a_2b}f^{a_3a_4b}
\bigg[
  (p_2^2+(p_3+p_4)^2)(p_3+p_4-p_2)_{\mu_1}g_{\mu_3\mu_4}
  +
  (p_2^2+p_3^2)p_3{}_{\mu_3}g_{\mu_1\mu_4}
}\\
  +
  (p_3+p_4-p_2)_{\mu_1}(p_3+p_1-p_2)_{\mu_4}p_3{}_{\mu_3}
\bigg]
+
\left(1\leftrightarrow 4\right).
\label{vertex:4D AAbjAj}
\end{multline}
Where $\left(1\leftrightarrow 4\right)$ means
the same expression exchanged all the index $1$ with $4$.

\section{Momentum Integration}
\label{appendix:calculations of momentum integration}
%

All the momentum integrals that we consider in this paper
are reduced to the general form of
\begin{equation}
I_{\mu_1\cdots\mu_n}(x_1,\ldots,x_N) \equiv
\int \frac{\mathrm{d}^4 k}
          {(2\pi)^4}
\frac{ k_{\mu_1} \cdots k_{\mu_n} }
     {\prod_{i=1}^{N}(k^2 + x_i )(q^2 + x_i )}.
\end{equation}
where $q = k - p$ and $p$ is the external momentum.
This expression is written as follows
using the Feynman parameterization and 
introducing $z$ which is a source of the momentum:
\begin{multline}
I_{\mu_1\cdots\mu_n}(x_i)
=
\int \frac{\mathrm{d}^4k }
          {(2\pi)^4}
\frac{\delta }{ \delta z_{\mu_1}}\cdots
\frac{\delta }{ \delta z_{\mu_n}}
\int_0^\infty \prod_{i=1}^N  \mathrm{d}\alpha_i  \mathrm{d}\beta_i
\\
\exp 
\left[-k^2\sum_{i=1}^N (\alpha_i+\beta_i)
 +k(2p\sum_{i=1}^N\beta_i + z)
 -\sum_{i=1}^N(\alpha_i+\beta_i)x_i
 -p^2\sum_{i=1}^N\beta_i
\right]_{z=0}.
\end{multline}
First we take the momentum integration with $k$ using the gaussian
and the differential with $z$.
%
Inserting an unity
$
1=\int_0^\infty \mathrm{d}\rho \delta
\left(\rho-\sum\alpha_i-\sum\beta_i\right)$,
and rescaling the parameters such as
$\alpha_i \rightarrow \rho\alpha_i$ and
$\beta_i \rightarrow \rho\beta_i$,
the expression is written as follows:
for odd $n=2m-1$
\begin{subequations}
\begin{multline}
I_{\mu_1\cdots\mu_{2m-1}}(x_i)
=
\sum_{k=1}^m
\frac{1}{ 2^{4+m-k}\pi^{2}}
\left[
 p_{\mu_1}\cdots p_{\mu_{2k-1}}
 \delta_{\mu_{2k}\mu_{2k+1}}\cdots
 \delta_{\mu_{2m-2}\mu_{2m-1}}
\right]_\mathrm{sym}\\
\times
\int_0^1\prod \mathrm{d}\alpha_i \mathrm{d}\beta_i
\left(\sum\beta_i\right)^{2k-1}
\delta\left(1 -\sum \alpha -\sum \beta\right)\\
\times
\int_0^\infty \mathrm{d}\rho\rho^l
\exp
\left[
 -\rho\left\{
       p^2\sum\alpha_i\sum\beta_i
       + \sum (\alpha_i+\beta_i)x_i
      \right\}
\right],
\label{eq:before rho int odd n}
\end{multline}
and for even $n=2m$
\begin{multline}
I_{\mu_1\cdots\mu_{2m}}(x_i)
=
\sum_{k=0}^m
\frac{1}{ 2^{4+m-k}\pi^{2}}
\left[
 p_{\mu_1}\cdots p_{\mu_{2k}}
 \delta_{\mu_{2k+1}\mu_{2k+2}}\cdots
 \delta_{\mu_{2m-1}\mu_{2m}}
\right]_\mathrm{sym}\\
\times
\int_0^1\prod  \mathrm{d}\alpha_i  \mathrm{d}\beta_i
\left(\sum\beta_i\right)^{2k}
\delta\left(1 -\sum \alpha -\sum \beta\right)\\
\times
\int_0^\infty \mathrm{d}\rho\rho^l
\exp
\left[
 -\rho\left\{
       p^2\sum\alpha_i\sum\beta_i
       + \sum (\alpha_i+\beta_i)x_i
      \right\}
\right],
\label{eq:before rho int even n}
\end{multline}
\label{eq:before rho int}
\end{subequations}
where $i$ runs from $0$ to $N$.
We define $l\equiv 2N+k-m-3$
and $[\qquad]_\mathrm{sym}$ as the symmetrization about indices like
$
\left[
\delta_{\mu_1 \mu_2} p_{\mu_3}
\right]_\mathrm{sym}
=\delta_{\mu_1 \mu_2} p_{\mu_3}
+\delta_{\mu_2 \mu_3} p_{\mu_1}
+\delta_{\mu_3 \mu_1} p_{\mu_2}.
$
We take $\sum\alpha_i+\sum\beta_i=1$ because of the 
$\delta$-function.

To calculate the expression,
we consider the $\rho$-integration
previous to the parameter integration of $\alpha_i$ and $\beta_i$.
First we consider the case when $l$ is positive value.
With an integration by parts we get,
\begin{multline}
\int_0^\infty \mathrm{d}\rho \rho^l
\exp
 \left[
  -\rho b \{aX + Y\}
 \right]\\
=
-\frac{\rho^l }{ b \{aX + Y\}}
 \mathrm{e}^{-\rho b \{aX + Y\}}
 \bigg|_0^\infty
+
 \int_0^\infty  \mathrm{d}\rho
 \frac{l \rho^{l-1} }{ b \{aX + Y\}}
 \mathrm{e}^{-\rho b \{aX + Y\}}.
\end{multline}
The first term of r.h.s.{} vanishes under the condition $b\{aX + Y\} > 0$.
Integrating by parts recursively,
we get the following result:
\begin{equation}
\int_0^\infty \mathrm{d}\rho \rho^l
\exp
 \left[
  -\rho b \{aX + Y\}
 \right]
=
\frac{(-1)^l }{ a^l b^{l+1}}\frac{\partial^l }
{ (\partial X)^l}
\left[\frac{1}{ aX+Y}\right]. 
\label{eq:differential equation}
\end{equation}
Going to the last expression,
we use the formula
\begin{equation}
\int_0^\infty \mathrm{d}\rho
\exp
 \left[
  -\rho b \{aX + Y\}
 \right]
=
\frac{1 }{ b\{aX + Y\} }.
\label{eq:integration of rho order 0}
\end{equation}

For negative $l$ using the relation
\begin{equation}
\frac{\partial^{|l|} }{ (\partial X)^{|l|}}
\int_0^\infty \frac{\mathrm{d}\rho }{ \rho^{|l|}}
\mathrm{e}^{-\rho b\{aX + Y\}}
=
(-ab)^{|l|}
\int_0^\infty \mathrm{d}\rho
\mathrm{e}^{-\rho b\{aX + Y\}}
\end{equation}
and \eqref{eq:integration of rho order 0},
we get a differential equation about $X$.
The differential equation is solved
by integrating with $X$ recursively,
and the solution is formally written by
\begin{multline}
\int_0^\infty \frac{\mathrm{d}\rho }{ \rho^{|l|}}
\mathrm{e}^{-\rho b\{aX + Y\}}
=
{(-1)^{|l|}  a^{|l|} b^{|l|-1}}
\bigg(
 \mathcal{C}_{|l|}+X\mathcal{C}_{|l|-1}+\cdots
\\
+
 \frac{X^{|l|-1}}{(|l|-1)!}\mathcal{C}_1+
 \frac{(aX-Y)^{|l|-1}}{(|l|-1)!}\frac{\ln (aX+Y)}{a^{|l|}}+\cdots
\bigg).
\label{eq:negative l formula}
\end{multline}
Here $\mathcal{C}_i$ arises from the $i$-th integral
and represented
using the parameter $\mathcal{X}$ which has the same dimension with $X$
as
\begin{align}
\mathcal{C}_1&=-\ln\mathcal{X},&
\mathcal{C}_2&=\mathcal{X},&
\mathcal{C}_3&=-\frac{\mathcal{X}^2}{4},&
\mathcal{C}_4&=\frac{\mathcal{X}^3}{18},&
\cdots.
\end{align}

Using the results,
we calculate the easiest formulae of $N=1$ and $x_i = 0$
taking $X=p^2$ as follows:
\begin{align}
&I(0) =
-\frac{1}{ 16\pi^2} (\ln{p^2}+\mathcal{C}_1),&
&I_{\mu}(0) =
-\frac{p_\mu}{ 32\pi^2} (\ln{p^2}+\mathcal{C}_1).
\label{eq:N=1,D=4,x=0,l=-1}
\end{align}
We can calculate the more general $n$ of this case
\begin{equation}
I_{\mu_1\cdots\mu_n}(0)\sim
\mathcal{C}_{\left[1+\frac{n}{2}\right]}+
p^2\mathcal{C}_{\left[\frac{n}{2}\right]}+
\cdots+
\frac{p^{2\left[\frac{n}{2}\right]}}{\left[\frac{n}{2}\right]!}
\mathcal{C}_{1}+\mathrm{const.}
\label{eq:N=1,D=4,x=0,general}
\end{equation}
where $[\quad]$ denotes the Gauss' notation
and $\mathcal{C}_i$ is represented
using the parameter $\mathcal{M}$ of mass dimension of one
by
\begin{align}
\mathcal{C}_1&=-\ln \mathcal{M}^2,&
\mathcal{C}_2&= \mathcal{M}^2,&
\mathcal{C}_3&=-\frac{\mathcal{M}^4}{4},&
\mathcal{C}_4&= \frac{\mathcal{M}^6}{18},&
\cdots.
\label{eq:estimation of constants}
\end{align}
The parameter $\mathcal{M}$ goes to $\infty$
just like the cut-off parameter.
Under this parametrization,
\eqref{eq:N=1,D=4,x=0,l=-1} is read
as the dimensionless divergence of
$\ln (\mathcal{M}/p)$.

\section{Infinite Sum}
\label{appendix:calculations of infinite sum}

Here we give a formula that we use the calculation of
an infinite sum of PV fields.
Under the Feynman gauge,
all the expressions of the diagrams listed in Figures%
~\ref{fig:loops} and \ref{fig:loops only Lambda4}
are reduced to the following forms;
\begin{equation}
\sum_{j=-\infty}^\infty
(-1)^j M_j^{2r}
\int \frac{\mathrm{d}^4k}{(2\pi)^4}
\frac{k_{\mu_1}\cdots k_{\mu_n}}{(k^2+M_j^2)^N(q^2+M_j^2)^N}.
\label{eq:general form of infinite sum}
\end{equation}
At the beginning, we consider $r=0$ case.
Inserting (\ref{eq:before rho int}) to this expression,
we find that the most important calculation is the summation with $j$
which is given after the $\rho$-integration,
\begin{multline}
\sum_{j=-\infty}^\infty
(-1)^j
\int_0^\infty \mathrm{d}\rho \rho^l
\exp
 \left[
  -\rho M^2b\{aX + |j|^2\}
 \right]\\
=\frac{(-1)^l  }{ a^{l} (M^2b)^{l+1}}
\frac{\partial^l }{ (\partial X)^l}
\left[
\frac{1}{ aX} - \frac{\pi^2 }{ 6} +\frac{7 \pi^4 }{ 360 } aX +
\cdots
\right],
\label{eq:rho integration and summation with r=0}
\end{multline}
where we take $M_j=M|j|$ and define $a$, $b$ and $X$ as follows:
\begin{align}
a&\equiv\frac{\sum_{i=1}^N\alpha_i \sum_{i=1}^N\beta_i}
        {\sum_{i=1}^N\alpha_i+\beta_i},&
b&\equiv\sum_{i=1}^N\alpha_i+\beta_i,&
X&\equiv\frac{p^2}{M^2}.
\end{align}
Going to the r.h.s.{} of \eqref{eq:rho integration and summation with r=0}
we use the formula
$
\sum_{j\in\mathbb{Z}}
 \frac{(-1)^j }{ A^2 + j^2}
 =\frac{\pi }{ A \sinh \pi A}$
and expand it.
The negative $l$ means multiple integral of order $l$
as mentioned Appendix~\ref{appendix:calculations of momentum integration}.
In the case of $l = -1$, for instance,
we get
\begin{multline}
\sum_{j=-\infty}^\infty
 (-1)^j
\int_0^\infty \frac{\mathrm{d}\rho}{\rho}
\exp
 \left[
  -\rho M^2b \{aX + |j|^2\}
 \right] \\
=
-\ln X +\frac{\pi^2}{6}aX - \frac{7\pi^4}{720}a^2X^2 -aC_1 +\cdots,
\end{multline}
where $C_1$ is a dimension less constant and determined by the relation
$
 \frac{\pi}{\sqrt{aX}\sinh \pi \sqrt{aX}}=
 \frac{\partial}{\partial X}
 \left(
   \frac{2}{a}\tanh \frac{\pi}{2}\sqrt{aX}
 \right)
$
such as
\begin{equation}
C_1=\frac{1}{a}
  \ln
    \left(
      \frac{\pi^2}{4}a
    \right).
\end{equation}
%

%
For $r = 1$
the key object is the following equation:
\begin{equation}
\sum_{j=-\infty}^\infty
 (-1)^j M_j^2
\int_0^\infty \mathrm{d}\rho \rho^l
\exp
 \left[
  -\rho M^2b \{aX + |j|^2\}
 \right].
\label{eq:rho integration order l with m2}
\end{equation}
Notice the relation
\begin{multline}
\frac{\partial}{\left(\partial M^2\right)}
\int_0^\infty \mathrm{d}\rho \rho^{l-1}
\exp
 \left[
  -\rho M^2b \{aX + |j|^2\}
 \right]\\
=
-b\{aX + |j|^2\}
\int_0^\infty \mathrm{d}\rho \rho^l
\exp
 \left[
  -\rho M^2b \{aX + |j|^2\}
 \right],
\label{eq:derivation with m2}
\end{multline}
and using \eqref{eq:rho integration and summation with r=0}
we get
\begin{multline}
\sum_{j=-\infty}^\infty
 (-1)^j M_j^2
\int_0^\infty \mathrm{d}\rho \rho^l
\exp
 \left[
  -\rho M^2b \{aX + |j|^2\}
 \right]\\
=
\frac{(-1)^{l+1}}{ a^{l-1}(M^2)^lb^{l+1}}
\left[
 X\left(\frac{\partial}{\partial X}\right)^l
 +l\left(\frac{\partial}{\partial X}\right)^{l-1}
\right]
\left[
 \frac{1}{ aX}-\frac{\pi^2}{ 6}+\frac{7\pi^2 }{ 360}aX+\cdots
\right].
\label{eq:rho integration and summation with r and m2}
\end{multline}
Using \eqref{eq:rho integration and summation with r=0} and
\eqref{eq:rho integration and summation with r and m2} iteratively,
we get the formula for $r \ge 1$,
\begin{multline}
\sum_{j=-\infty}^\infty
 (-1)^j M_j^{2r}
\int_0^\infty \mathrm{d}\rho \rho^l
\exp
 \left[
  -\rho M^2b \{aX + |j|^2\}
 \right]\\
=
\frac{(-1)^{l+r}}{ a^{l-r}(M^2)^{l+1-r}b^{l+1}}
\sum_{q=0}^r
\begin{pmatrix}r\\q\end{pmatrix}
\frac{l!}{(l-q)!}X^{r-q}
\left(\frac{\partial}{\partial X}\right)^{l-q}
\\
\left[
 \frac{1}{ aX}-\frac{\pi^2}{ 6}+\frac{7\pi^2 }{ 360}aX+\cdots
\right].
\label{eq:general formula of rho integration and summation}
\end{multline}

Here we give explicit calculations
which we use in Section~\ref{sec:one-loop contributions}%
~and~\ref{sec:one-loop contributions in finite}.
For $\Lambda^0$ order, all the integrals are given by
\eqref{eq:general form of infinite sum} with $N=1$,
we have to calculate up to $n=2$ with $r=0$
and $n=0$ with $r=2$.
Since these integrals give the $\rho$-integration with $l=-1$ and $l=-2$
there appear dimension less constants $C_1$ and $C_2$ as follows:
\begin{subequations}
\begin{align}
&
\sum_{j=-\infty}^\infty
(-1)^j I (M_j^2)= \frac{1}{ 16 \pi^2}
\left[
 -2 \ln\left(\frac{\pi p}{ 2M}\right)
 +2+\frac{\pi^2p^2}{ 36M^2}
 +O\left(M^{-4}\right)
\right],
\\ \nonumber \\
&
\sum_{j=-\infty}^\infty
 (-1)^j I_{\mu_1} (M_j^2) =
\frac{p_{\mu_1} }{ 16 \pi^2}
\left[
 -\ln\left(\frac{\pi p}{ 2M}\right)
 +1+\frac{\pi^2p^2}{ 72M^2}
 +O\left(M^{-4}\right)
\right],
\\ \nonumber \\
&
\sum_{j=-\infty}^\infty
 (-1)^j I_{\mu_1\mu_2} (M_j^2)
=
\frac{p_{\mu_1}p_{\mu_2}}{ 16\pi^2}
\left[
 -\frac{2}{ 3} \ln\left(\frac{\pi p }{ 2 M }\right)
 +\frac{13}{ 18} +\frac{\pi^2 p^2 }{ 120 M^2}
 +O\left(M^{-4}\right)
\right]\nonumber \\
&\qquad
+
\frac{p^2 \delta_{\mu_1 \mu_2} }{ 32 \pi^2}
\left[
 \frac{M^2 }{ 30p^2}C_2 
 +\frac{1}{ 3} \ln\left(\frac{\pi p }{ 2 M }\right)
 -\frac{4}{ 9}-\frac{\pi^2 p^2 }{ 360 M^2}
 +O\left(M^{-4}\right)
\right],
\\ \nonumber \\
&
\sum_{j=-\infty}^\infty
 (-1)^j M_j^2 I (M_j^2)
= 
\frac{1}{ 16\pi^2}
\left[
-\frac{M^2}{ 30}C_2 +  \frac{p^2}{ 6} 
 - \frac{\pi^2 p^4 }{ 360M^2} +O\left(M^{-4}\right)
\right].
\end{align}
\label{eq:result of infinite sum}
\end{subequations}
Similarly for $\Lambda^{-4}$ order
all the integrals are given by $N=2$.
The most divergent parts are calculated
\begin{subequations}
\begin{align}
&
\sum_{j=-\infty}^\infty
 (-1)^j
I_{\mu_1\cdots\mu_{10}} (M_j^2, M_j^2)
=
\frac{
      \left[
       \delta_{\mu_1\mu_2}\cdots\delta_{\mu_9\mu_{10}}
      \right]_\mathrm{sym}
     }
{ 2^9 \pi^2}
\frac{M^6}{2772}C_4 + O\left(M^4\right),
\\
&
\sum_{j=-\infty}^\infty
 (-1)^j M_j^2
I_{\mu_1\cdots\mu_{8}} (M_j^2, M_j^2)
=
-\frac{
      \left[
       \delta_{\mu_1\mu_2}\cdots\delta_{\mu_7\mu_8}
      \right]_\mathrm{sym}}
{ 2^8 \pi^2}
\frac{3M^6}{2772}C_4 + O\left(M^4\right),
\\
&
\sum_{j=-\infty}^\infty
 (-1)^j M_j^4
I_{\mu_1\cdots\mu_{6}} (M_j^2, M_j^2)
=
\frac{
      \left[
       \delta_{\mu_1\mu_2}\cdots\delta_{\mu_5\mu_6}
      \right]_\mathrm{sym}}
{ 2^7 \pi^2}
\frac{6M^6}{2772}C_4 + O\left(M^4\right),
\\
&
\sum_{j=-\infty}^\infty
 (-1)^j M_j^6
I_{\mu_1\cdots\mu_{4}} (M_j^2, M_j^2)
=
-\frac{
      \left[
       \delta_{\mu_1\mu_2}\delta_{\mu_3\mu_4}
      \right]_\mathrm{sym}}
{ 2^6 \pi^2}
\frac{6M^6}{2772}C_4 + O\left(M^4\right),
\\
&
\sum_{j=-\infty}^\infty
 (-1)^j M_j^8
I_{\mu_1\mu_{2}} (M_j^2, M_j^2)
=
O\left(M^4\right),
\end{align}
\label{eq:result of infinite sum N=2}
\end{subequations}
where $C_4$ is the integration constant
arising from the fourth order integration of $X$
which is caused by the $\rho$-integration with $l=-4$.

%
%
\providecommand{\href}[2]{#2}\begingroup\raggedright\endgroup


\begin{thebibliography}{10}

\bibitem{Frolov-Slavnov93}
S.~A. Frolov and A.~A. Slavnov, ``{An Invariant Regularization of the Standard
  Model},'' {\em Phys. Lett.} {\bf B309} (1993)
344--350.

\bibitem{Frolov-Slavnov94}
S.~A. Frolov and A.~A. Slavnov, ``{Removing Fermion Doublers in Chiral Gauge
  Theories on the Lattice},'' {\em Nucl. Phys.} {\bf B411} (1994) 647--664,
\href{http://www.arXiv.org/abs/hep-lat/9303004}{{\tt hep-lat/9303004}}.

\bibitem{Nittoh-Ebihara98MPLA}
K.~Nittoh and T.~Ebihara, ``{Regularization ambiguity problem for the
  Chern-Simons shift},'' {\em Mod. Phys. Lett.} {\bf A13} (1998) 2231--2237,
\href{http://www.arXiv.org/abs/hep-th/9809036}{{\tt hep-th/9809036}}.

\bibitem{Nittoh99phd}
K.~Nittoh, {\em A New Regularization Scheme with Modified Pauli-Villars Fields
  and Its Application}.
\newblock {Ph.D.} dissertation, Chiba University, 1999.

\bibitem{Jack-Jones97}
I.~Jack and D.~Jones, ``{Regularization of Supersymmetric Theories},'' in {\em
  Perspectives on Supersymmetry}, G.~Kane, ed.
\newblock World Scientific, 1997.
\newblock \href{http://www.arXiv.org/abs/hep-ph/9707278}{{\tt hep-ph/9707278}}.

\bibitem{Kao-Lee-Lee95}
H.-C. Kao, K.~Lee, and T.~Lee, ``{The Chern-Simons Coefficient in
  Supersymmetric Yang-Mills Chern-Simons Theories},'' {\em Phys. Lett.} {\bf
  B373} (1996) 94, \href{http://www.arXiv.org/abs/hep-th/9506170}{{\tt
  hep-th/9506170}}.

\bibitem{Koiwa98}
Y.~Koiwa, ``{Chern-Simons Shift in Super Gauge Theory},'' October, 1998.
\newblock Talk on JPS meeting, Akita University, Akita.

\bibitem{Slavnov72}
A.~Slavnov, ``{Invariant Regularization of Gauge Theories},'' {\em Theor. Math.
  Phys.} {\bf 13} (1972) 1064.

\bibitem{Slavnov77}
A.~Slavnov, ``{Pauli-Villars Regularization for Non-Abelian Gauge Theories},''
  {\em Theor. Math. Phys.} {\bf 33} (1977) 210.

\bibitem{Faddeev-Slavnov91}
L.~Faddeev and A.~Slavnov, {\em {Gauge Field, Introduction to Quantum Theory}}.
\newblock Addison-Wesley, Redwood, 2nd~ed., 1991.

\bibitem{Martin-Ruiz95}
C.~Martin and F.~Ruiz~Ruiz, ``{Higher covariant derivative Pauli-Villars
  regularization does not lead to a consistent QCD},'' {\em Nucl. Phys.} {\bf
  B436} (1995) 545.

\bibitem{Leon-Martin-Ruiz95}
J.~Leon, C.~Martin, and F.~Ruiz~Ruiz, ``{Unitarity violation in non-abelian
  Pauli-Villars regularization},'' {\em Phys. Lett.} {\bf B355} (1995) 531.

\bibitem{Martin-Ruiz95PL}
C.~Martin and F.~Ruiz~Ruiz, ``{Higher covariant derivative regulators and
  non-multiplicative renormalization},'' {\em Phys. Lett.} {\bf B343} (1995)
  218.

\bibitem{Asorey-Falceto95}
M.~Asorey and F.~Falceto, ``{On the Consistency of the Regularization of Gauge
  Theories by High Covariant Derivatives},'' {\em Phys. Rev.} {\bf D54} (1996)
  5290, \href{http://www.arXiv.org/abs/hep-th/9502025}{{\tt hep-th/9502025}}.

\bibitem{Pronin-Stepanyantz97PL}
P.~Pronin and K.~Stepanyants, ``{One-loop counterterms for higher derivative
  regularized Lagrangians},'' {\em Phys. Lett.} {\bf B414} (1997) 117,
  \href{http://www.arXiv.org/abs/hep-th/9707008}{{\tt hep-th/9707008}}.

\bibitem{NittohEP118}
K.~Nittoh, ``{A Note on the Quadratic Divergence in Hybrid Regularization},''
  {\em Mod. Phys. Lett.} {\bf A15} (2000) 955,
\href{http://www.arXiv.org/abs/hep-th/0005175}{{\tt hep-th/0005175}}.

\bibitem{Pronin-Stepanyantz97NP}
P.~Pronin and K.~Stepanyants, ``{One-loop counterterms for the dimensional
  regularization of arbitrary Lagrangians},'' {\em Nucl. Phys.} {\bf B485}
  (1997) 517, \href{http://www.arXiv.org/abs/hep-th/9605206}{{\tt
  hep-th/9605206}}.

\bibitem{Politzer}
H.~D. Politzer, ``{Reliable Perturbative Results for Strong Interactions?},''
  {\em Phys. Rev. Lett.} {\bf 30} (June, 1973) 1346.

\bibitem{Gross-Wilczek73PRL}
D.~J. Gross and F.~Wilczek, ``{Ultraviolet Behavior of Non-Abelian Gauge
  Theories},'' {\em Phys. Rev. Lett.} {\bf 30} (1973)
1343--1346.

\bibitem{Gross-Wilczek73PhysRev}
D.~J. Gross and F.~Wilczek, ``{Asymptotically Free Gauge Theories. I},'' {\em
  Phys. Rev.} {\bf D8} (1973)
3633.

\bibitem{Bakeyev-Slavnov96}
T.~D. Bakeyev and A.~A. Slavnov, ``Higher covariant derivative regularization
  revisited,'' {\em Mod. Phys. Lett.} {\bf A11} (1996) 1539--1554,
\href{http://www.arXiv.org/abs/hep-th/9601092}{{\tt hep-th/9601092}}.

\bibitem{Warr87_1}
B.~J. Warr, ``{Renormalization of Gauge Theories Using Effective Lagrangians,
  I},'' {\em Ann. Phys.} {\bf 183} (1988) 1.

\end{thebibliography}
\end{document}